\documentclass[10pt,jcp,aip,twocolumn,showpacs,longbibliography,superscriptaddress]{revtex4-1}
\usepackage{epsfig,amsmath,amssymb,graphicx,color,calc,epstopdf}
\usepackage{color}
\usepackage{inputenc,bm}
\usepackage{hyperref}
\usepackage[mathscr]{euscript}

\definecolor{gris}{gray}{0.8}
\definecolor{grisbleu}{rgb}{0.47,0.6,0.82}

\newcommand{\hide}[1]{&{\textcolor{gris}{{#1}}}\\}
\newcommand{\killinpaper}[1]{\textcolor{grisbleu}{{#1}}}
\renewcommand{\hide}[1]{}
\renewcommand{\killinpaper}[1]{}

\def\PP{{\cal P}}

\def\RR{{\cal R}}\def\LL{{\cal L}}  
 
 \def\QQ{{\cal Q}}

\newcommand{\ttt}{\mathfrak{t}}
 
\newcommand{\RRi}{\mathcal{S}}  
\newcommand{\bb}{\wp} 

\def\to{\rightarrow}

\newcommand{\Out}{{\mathrm{out}}}
\newcommand{\In}{{\mathrm{in}}}
\newcommand{\grad}{\mathbf{grad}}
\newcommand{\curl}{\mathbf{curl}}
\renewcommand{\div}{\mathrm{div}}
\newcommand{\stick}{\mathrm{stick}}
\newcommand{\slip}{\mathrm{slip}}
\newcommand{\AAA}{\mathcal{A}}

\newcommand{\SO}{\mathrm{SO}}
\newcommand{\SSS}{\mathbb{S}}
\newcommand{\vect}[1]{{\begin{pmatrix}#1\end{pmatrix}}}
\newcommand{\diag}{\mathrm{diag}}

\newcommand{\dd}{\mathrm{d}}
\newcommand{\p}{\partial}

\newcommand{\scp}[1]{\left\langle #1\right\rangle} 
\newcommand{\scpb}[1]{\left[ #1\right]} 
\renewcommand{\vec}{\mathbf}
\newcommand{\hvec}[1]{\hat{\vec{#1}}}
\newcommand{\tvec}[1]{\tilde{\vec{#1}}}

\newcommand{\cst}{c}
\newcommand{\prefix}{\alpha}
\newcommand{\factorout}{\alpha_2}
\newcommand{\coeff}{\alpha}

\newcommand{\etashear }{\eta_{\mathrm{S}}}
\newcommand{\etabulk }{\eta_{\mathrm{B}}}

\newcommand{\PF}{\dot{\bm{\mathcal{F}}}}


\usepackage{xr}

\begin{document}
\title{A Microscopic Model of the Stokes--Einstein Relation in Arbitrary Dimension}
\author{Benoit Charbonneau}
\affiliation{Department of Pure Mathematics, University of Waterloo, Waterloo, Ontario N2L 3G3, Canada}
\affiliation{Department of Physics and Astronomy, University of Waterloo, Waterloo, Ontario N2L 3G3, Canada}
\author{Patrick Charbonneau}
\affiliation{Department of Chemistry, Duke University, Durham,
	North Carolina 27708, USA}
\affiliation{Department of Physics, Duke University, Durham,
	North Carolina 27708, USA}
\author{Grzegorz Szamel}
\affiliation{Department of Chemistry, Colorado State University, Fort Collins, CO 80523, USA}

\begin{abstract}
The Stokes--Einstein relation (SER) is one of most robust and widely employed results from the theory of liquids. Yet sizable deviations can be observed for self-solvation, which cannot be explained by the standard hydrodynamic derivation. Here, we revisit the work of Masters and Madden [J.~Chem.~Phys.~\textbf{74}, 2450-2459 (1981)], who first solved a statistical mechanics model of the SER using the projection operator formalism. By generalizing their analysis to all spatial dimensions and to partially structured solvents, we identify a potential microscopic origin of some of these deviations. We also reproduce the SER result from the exact dynamics of infinite-dimensional fluids.
\end{abstract}

\pacs{}

\date{\today}
\maketitle

\section{Introduction}

Einstein's \emph{annus mirabilis} saw the resolution of three 
problems at the interfaces between mechanics, thermodynamics and electrodynamics~\cite{Renn:2005}: the photoelectric effect, black-body radiation and Brownian motion.
Of the three, the last most strongly bears the imprint of its liminal origin. Like his contemporaries Smoluchowski and Sutherland, Einstein 
directly borrowed Stokes' hydrodynamic result for the drag on a hard particle in a continuum fluid to obtain the result now commonly known as the Stokes--Einstein relation (SER). Despite the obvious idealization this implied~\cite{Nye:1972}, the ensuing relation between the diffusivity of a particle $D$, its (hydrodynamic) radius $\AAA$, and the solvent shear viscosity $\eta_\mathrm{S}$ at inverse temperature $\beta$,
\begin{equation}
\beta D\eta_\mathrm{S}=
\begin{cases}
\frac{1}{6\pi \AAA},&\mathrm{stick},\\
\frac{1}{4\pi \AAA},&\mathrm{slip},
\end{cases}
\end{equation}
for either stick or slip solvent boundary conditions, is remarkable. It provides (nearly) quantitative predictions for the diffusion of spherical particles ranging from the molecular to the colloidal scale~\cite{Boon:1980,Bian:2016}, and qualitative estimates for an even broader range of systems. In this context, it is not surprising that it took the better part of the following century for a first fully microscopic derivation of SER even to be sought out. 

Masters and Madden (MM) were the first to accomplish this technical feat~\cite{Masters:1981,Keyes:1985,footnote1}. In order to reproduce the classical result without borrowing from continuum hydrodynamics, MM used 
tools of equilibrium and out-of-equilibrium statistical mechanics that simply did not exist at the start of the 20th century. Most notably, they employed the projection operator formalism that had grown out of the study of liquids in the preceding decades and given rise to various mode-coupling theories~\cite{Hansen:1986,Zwanzig:2001} 
(and eventually to a description of supercooled liquids~\cite{Gotze:2009}). By applying this formalism, MM shed some light on some of the less controlled aspects of SER and concluded a decade-long discussion triggered by the derivation by Keyes and Oppenheim of a result intermediate between the results for stick and slip boundary conditions~\cite{Keyes:1973}. MM's work, however, has since had but limited impact on the experimental study of SER. Their treatment was as elegant as their end result was uncontroversial. 

Careful numerical simulations have recently uncovered systematic deviations from SER that bring renewed interest in considering a fully microscopic model of diffusion. More specifically,  it has been shown that diffusion is systematically enhanced with respect to SER predictions when the radius, $A$, of a solvated hard sphere becomes comparable to the radius, $a$, of the solvating hard spheres~\cite{Schmidt:2003,Charbonneau:2013,Bian:2016}. At first glance, the existence of such a deviation might not seem surprising. Although Stokes' expression is exact for a large particle in a continuum fluid, the treatment must clearly break down once the granularity of the solvent becomes comparable to that of the solute. It is instead astonishing that SER could hold at all on such scales. Yet the bigger surprise is the sign of the deviation. Once the proper boundary condition is chosen~\cite{Schmidt:2003,Charbonneau:2013}, the correction due to the finiteness of the Schmidt number, which has been presumed dominant in real systems, indeed goes in the direction opposite to what is observed in simulations~\cite{BalboaUsabiaga:2013}. A larger source of discrepancy than this effect must therefore have been overlooked.

A somewhat different set of advances also motivate revisiting MM's analysis. Although on a different scale than in the early 1900s, a borderline problem involving
SER has emerged once again. Research on glasses, which has recently progressed most quickly at the interface of the infinite-dimensional ($d=\infty$) mean-field description of disordered systems and standard finite-$d$ liquid state theory~\cite{Charbonneau:2017}, has uncovered a number of discrepancies between the two approaches. Most significantly, the exact dynamics from the former predicts a dynamical arrest at densities infinitely smaller than the latter for hard sphere glasses~\cite{Ikeda:2010,Schmid:2010,Charbonneau:2011,Maimbourg:2016,Charbonneau:2017}. (A similar effect is also observed in random Lorentz gases~\cite{Jin:2015}.) In addition, as $d$ increases, corrections to SER for self-diffusion systematically grow with fluid density~\cite{Charbonneau:2013,Maimbourg:2016,Charbonneau:2017}, independently of the \emph{breakdown} of SER observed in deeply supercooled liquids~\cite{Fujara:1992,Cicerone:1993,Stillinger:1994,Tarjus:1995,Cicerone:1996,Chang:1997,
Perera:1998,Xia:2001,Debenedetti:2001,Kumar:2006, Charbonneau:2013,Charbonneau:2014}. Considering the high-dimensional behavior of a microscopic model of SER might thus provide much needed insight into these observations.

In this article, we present a generalized and refined treatment of the projection operator approach developed by MM. This alternate derivation allows us to solve the model they considered in all $d$. Its validation against the generalized SER obtained from hydrodynamics then provides a number of microscopic insight into the theory of the liquid state. The plan for the rest of this paper is as follows. In Sec.~\ref{sec:static}, we define the minimal microscopic model of a Brownian particle and describe the elementary observables that characterize its static properties. In Sec.~\ref{sec:particle}, we write the theory for the dynamics of the Brownian particle, and in Sec.~\ref{sec:solvent} for that of the solvent. In Secs.~\ref{sec:unphysical} and \ref{sec:physical}, we then solve this theory for different models of the solvent structure. Section~\ref{sec:conclusion} concludes with a summary of the results and of possible extensions. 
Throughout the text, we have tried to follow closely the notation of MM, but deviations are necessary to disambiguate some of the expressions.

\section{Model description and static quantities}
\label{sec:static}
As a minimal microscopic model of the SER, we consider a binary mixture consisting of a single hard sphere (the solute) of mass $M$ and radius $A$ solvated in a fluid of $N_\mathrm{f}\gg 1$ (much) smaller hard spheres of mass $m$ and radius $a$. Note that because we expect the solute to undergo Brownian motion under the influence of the solvent, we refer to is as 
the Brownian particle. Note also that for notational convenience, we distinguish quantities associated with the Brownian particle from those associated with the solvent by using uppercase letters for the former and lowercase letters for the latter. 

The Hamiltonian for the fluid particles at positions $\mathbf{r}$ and with momenta $\mathbf{p}$ and the Brownian particle 
at $\mathbf{R}$ with momentum $\mathbf{P}$ is then
\begin{equation}\label{Ham}
\mathcal{H}=\sum_{i=1}^{N_\mathrm{f}} \frac{|\mathbf{p}|^2}{2m}+\frac{|\mathbf{P}|^2}{2M}+\sum_{i>j=1}^{N_\mathrm{f}}u(|\mathbf{r}_i-\mathbf{r}_j|)+\sum_{i=1}^{N_\mathrm{f}} U(|\mathbf{r}_i-\mathbf{R}|),
\end{equation}
where, in order to avoid the notational complications associated with discontinuous interaction potentials, we replace the hard interactions by harshly repulsive ones
\begin{align}\nonumber
u(x)&=\begin{cases}
|x-2a|^\mu, & x<2a,\\
0,&x\geq 2a,
\end{cases}\\
U(x)&=\begin{cases}
|x-\AAA|^\mu, & x<\AAA,\\
0,&x\geq \AAA,
\end{cases}
\end{align}
with an exponent $\mu$ that is arbitrarily large and $\AAA=A+a$. The hard sphere limit, $\mu\rightarrow\infty$, is however implicitly taken throughout the calculation. 

The particles evolve with Newtonian dynamics within a cubic box of volume $V$ under periodic boundary condition, but the model is ultimately solved in the thermodynamic limit, $N_\mathrm{f}\rightarrow\infty$ and $V\rightarrow\infty$, keeping the solvent number density, $N_\mathrm{f}/V\rightarrow\rho$, asymptotically constant. The SER should thus be recovered after taking the limit $a/A\rightarrow 0$, with particle masses proportional to their volume, i.e., $m/M\sim (a/A)^{d}$. 

The structure of the system can be characterized in (left) real and (right) reciprocal space by its instantaneous densities:
\begin{align*}
N(\mathbf{R})&=\delta(\mathbf{r}-\mathbf{R}),&N(\mathbf{K})&=e^{i\mathbf{K}\cdot\mathbf{R}};\\
n(\mathbf{r})&=\sum_{i=1}^{N_\mathrm{f}}\delta(\mathbf{r}-\mathbf{r}_i),&n(\mathbf{K})&=\sum_{i=1}^{N_\mathrm{f}}e^{i\mathbf{K}\cdot\mathbf{r}_i};\\
\tilde{n}(\mathbf{r})&=\sum_{i=1}^{N_\mathrm{f}}\delta[\mathbf{r}-(\mathbf{r}_i-\mathbf{R})],&\tilde{n}(\mathbf{K})&=n(\mathbf{K})N(-\mathbf{K});\\
\mathbf{p}(\mathbf{r})&=\sum_{i=1}^{N_\mathrm{f}}\mathbf{p}_i\delta(\mathbf{r}-\mathbf{r}_i),&\vec{p}(\vec{K})&=\sum_{i=1}^{N_\mathrm{f}}\vec{p}_ie^{i\vec{K}\cdot\vec{r}_i};\\
\tilde{\mathbf{p}}(\mathbf{r})&=\sum_{i=1}^{N_\mathrm{f}}\mathbf{p}_i\delta[\mathbf{r}-(\mathbf{r}_i-\vec{R})],&\tilde{\mathbf{p}}(\mathbf{K})&=\mathbf{p}(\mathbf{K})N(-\mathbf{K}).
\end{align*}
Note the subtle difference between $n$ and $\tilde{n}$ (and between $\vec{p}$ and $\tilde{\vec{p}}$). The former is absolute, while the latter is defined in the reference frame of the Brownian particle. 
Upon averaging we get the standard radial distribution functions and structure factors of a binary system
\begin{align*}
\rho G(r)&=\langle\tilde{n}(\mathbf{r)}\rangle,\\
S(\mathbf{K})&=1+\rho\int_V \dd \mathbf{r}[G(r)-1]e^{i\mathbf{K}\cdot\mathbf{r}}\\
&=1+\langle\tilde{n}(\mathbf{K)}\rangle-N_\mathrm{f}\delta_{\vec{K},\vec{0}},\\
\rho g(r)&=\frac{1}{N_\mathrm{f}}\left\langle \sum_{i\neq j}\delta[\mathbf{r}-(\mathbf{r}_i-\mathbf{r}_j)]\right\rangle,\\
s(\mathbf{K})&=1+\rho\int_V \dd\mathbf{r} [g(r)-1] e^{i\mathbf{K}\cdot\mathbf{r}}\\
&=\frac{1}{N_\mathrm{f}}\langle n(\mathbf{K})n(-\mathbf{K})\rangle-N_\mathrm{f}\delta_{\vec{K},\vec{0}}.
\end{align*}

Two other static quantities of interest are the mean-square momentum of the Brownian particle, $\langle \mathbf{P}\cdot\mathbf{P}\rangle=dM/\beta$,
and the mean-square force exerted by the solvent on that particle,
\begin{equation}\label{eq:aveF}
\beta\langle \mathbf{F}\cdot\mathbf{F}\rangle=\rho\int_V\dd\mathbf{r} G(r)\nabla\cdot\nabla U(\mathbf{r}),
\end{equation}
where $\mathbf{F} = - \partial_{\mathbf{R}} \sum_{i=1}^{N_\mathrm{f}} U(|\mathbf{R}-\mathbf{r}_i|)$ is the instantaneous force exerted by the solvent on the particle. Note that the integral of the trace of the Hessian, $\nabla\nabla U(r)$, on right-hand side of Eq.~\eqref{eq:aveF} is obtained using the Yvon theorem~\cite{Hansen:1986}. For a spherically symmetric contact potential, as we consider here, the Hessian then takes the generic form
\begin{equation}
\label{eq:sphericalhessian}
\nabla\nabla U(r)=\frac{\delta(r-\AAA)}{S_{d-1}r^{d-1}\rho G(r)}(\phi \mathbf{I}+\psi \hat{\mathbf{r}}\otimes\hat{\mathbf{r}}),
\end{equation}
where $\mathbf{I}$ is the identity matrix, $\hat{\mathbf{r}}$ is a unit vector, and  $S_{d-1}$ is the surface area of a $d$-dimensional ball. In other words, $S_d$ is the volume of the $d$-dimensional sphere $\SSS^{d}$. The right-hand-side of Eq.~\eqref{eq:sphericalhessian} also defines two scalar quantities,
$\phi$ and $\psi$, with which we express the force as
\begin{equation}\label{forcecorrst}
\beta\langle \mathbf{F}\cdot\mathbf{F}\rangle=d\phi+\psi.
\end{equation}
Physically-speaking, $\phi$ and $\psi$ determine how momentum is exchanged between the solvent and the Brownian particle. Pure slip, which corresponds to the solvent forcing only radially on the particle and therefore allows no tangential transfer of momentum, microscopically corresponds to $\phi=0$; pure stick, which allows forcing from all components of the solvent collisions, corresponds to $\psi=0$. Intermediate cases with a finite ratio $\phi/\psi$ are also conceivable, but are not considered here.
Note that while this procedure to impose the stick boundary condition for the fluid velocity relies upon an intuitively justified mathematical model, microscopically the phenomenon in fact results from the existence of surface roughness. Note also that it is unclear whether the present model can reproduce proper boundary conditions for a fluid temperature field. However, this concern is not significant for the present study, because we assume that the system is fully thermalized at all times.

\section{Particle Dynamics}
\label{sec:particle}
In this section we define the frequency-dependent diffusion coefficient and express it in terms of a friction coefficient, $\xi$, using the projection operator formalism. We also obtain a formal expression for the friction
coefficient in terms of correlation functions of the fluctuations of the solvent momenta, $l(\vec{K})$ and $t(\vec{K})$, that are, respectively, longitudinal and transverse in the reference frame of the Brownian particle.

The frequency-dependent diffusion coefficient of the Brownian particle is defined as the Laplace transform of its velocity autocorrelation function 
\begin{align}\label{Domega}
D(\omega)&=\frac{1}{dM^2}\int_0^\infty \dd t e^{i\omega t}\langle \textbf{P}(t)\cdot\textbf{P}\rangle\\
&=\frac{1}{dM^2}\langle \mathbf{P}(\omega)\cdot\mathbf{P}\rangle.
\end{align}
Using this definition allows us to naturally separate the timescales of the various physical processes at play. In particular, the heavy Brownian 
particle moves much slower (on average) than the light solvent particles, and the average of the frictional 
force that the solvent exerts on the particle also evolves slowly.
It is thus natural to follow the approach of MM, 
and to select the momentum
of the Brownian particle and the force acting on it as slow variables. We then 
use the projector operator formalism to
analyze the evolution of the remaining variables. 

Explicitly, we define the projection operator $\PP$ as applied on an arbitrary vector $\vec{B}$, as
\begin{equation}\label{Pproj}
\PP\vec{B} = \frac{\scp{\vec{B}\cdot \vec{P}}}{\scp{\vec{P}\cdot \vec{P}}}\vec{P}+\frac{\scp{\vec{B}\cdot \vec{F}}}{\scp{\vec{F}\cdot \vec{F}}}\vec{F},
\end{equation}
and the orthogonal projection operator $\QQ = 1 - \PP$. Using standard manipulations\cite{Hansen:1986}, 
it is then possible to rewrite Eq.~\eqref{Domega} as 
\begin{equation}\label{proj1}
\beta D(\omega) = \left[\displaystyle-iM\omega+ M\xi(\omega)\right]^{-1},
\end{equation}
with a frequency-dependent friction coefficient
\begin{equation}\label{friction}
\xi(\omega) = \frac{\scp{\vec{F}\cdot\vec{F}}}{\displaystyle \scp{\vec{P}\cdot\vec{P}}\Bigl(-i\omega +\frac{\scp{\PF(\omega)\cdot \PF}}{\scp{\vec{F}\cdot \vec{F}}}\Bigr)}.
\end{equation}
Here, $\PF(\omega)$ is the Laplace transform of $\PF(t)$, which is the projected time derivative 
of the force, $\QQ \dot{\vec{F}}$, evolving with the so-called projected dynamics, \textit{i.e.} with an evolution operator
from which the slow variables have been projected out,
\begin{equation}\label{projecttimederF}
\PF(t) = e^{i\QQ\LL t}\QQ\dot{\vec{F}}
=e^{i\QQ\LL t}i\QQ \LL{\vec{F}}.
\end{equation}
The Liouville operator\cite{Hansen:1986}, $\LL$, is obtained from the Poisson bracket with the Hamiltonian defined in Eq.~\eqref{Ham}, that is
\begin{equation}\label{Liouville}
\LL B = i \{ \mathcal{H}, B \},
\end{equation}
which makes $\QQ\LL$ the projected Liouville operator. 

Using Eq.~\eqref{forcecorrst}, we thus have
\begin{equation}
M\xi(\omega;\phi,\psi)=\frac{\phi+\psi/d}{\left[-i\omega+\frac{\beta\langle\PF(\omega)\cdot\PF\rangle}{d\phi+\psi}\right]},
\end{equation}
which at zero frequency gives 
\begin{align}
\label{eq:mxiboundary}
M\xi(0;0,\psi)&=\frac{\psi^2}{d\beta\langle\PF(0)\cdot\PF\rangle}\\
M\xi(0;\phi,0)&=\frac{d\phi^2}{\beta\langle\PF(0)\cdot\PF\rangle}\nonumber
\end{align}
for slip and stick boundary conditions, respectively.

Denoting $\PF$ the variable $\PF(t)$ at initial time, $t=0$, we can use Eq.~\eqref{Pproj} to write
\begin{align}\label{X0a}
\PF =\QQ \dot{\vec{F}} \notag  =& \int \dd\vec{r}\left[\frac{\tilde{\vec{p}}(\vec{r})}{m}-\frac{\vec{P}}{M}\tilde{n}(\vec{r})\right]\cdot \nabla\nabla U(\vec{r}) \notag \\ 
&  -\left[\frac{\rho}{dM}\int \dd\vec{r}G(\vec{r})\nabla\cdot \nabla U(\vec{r})\right]\vec{P},
\end{align}
where the second term originates directly from the projection. Using the spherical symmetry of the
interaction potential we can then rewrite Eq.~\eqref{X0a} as
\begin{equation}\label{X0}
\PF=\int \dd\mathbf{r}\left\{\frac{\tilde{\mathbf{p}}(\mathbf{r})}{m}-\frac{\mathbf{P}}{M}[\tilde{n}(\mathbf{r})-\rho G(r)]\right\}\cdot\nabla\nabla U(r).
\end{equation}
As argued by MM, the 
second term within the curly brackets contributes a factor of order $\left(m/M\right)^{1/2}$ (or, equivalently, $(a/A)^{d/2}$) less than the first. Hence, if the Brownian particle is much more massive that the solvent particles, we can justifiably drop it. Note, however, that this argument is only valid because applying the projection operator subtracts the average fluid density around the particle. After this subtraction, the second term is a product of a slowly evolving quantity, $\vec{P}$ and a rapidly evolving fluctuation of the fluid density around the particle, $\tilde{n} - \rho G(r)$. The time integral of the square of the second term is therefore determined by that of $\langle[\tilde{n} - \rho G(r)]^2\rangle$, because $\vec{P}$ is approximately constant on the time scale on which this quantity decays.

Following these approximations, only the fluctuations of the force 
exerted by the solvent on the \emph{immobile} Brownian particle remain. 
For a spherically symmetric contact potential, using Eq.~\eqref{eq:sphericalhessian} gives 
\begin{align}
\PF &\approx 
\int \dd\mathbf{r}\frac{\tilde{\mathbf{p}}(\mathbf{r})}{m}\cdot\nabla\nabla U(\mathbf{r}) \\
&=\int \dd\mathbf{r}\frac{\tilde{\mathbf{p}}(\mathbf{r})}{m \rho S_{d-1}r^{d-1} G(r)}\delta(r-\AAA)(\phi \mathbf{I}+\psi \hat{\mathbf{r}}\otimes\hat{\mathbf{r}})\nonumber.
\end{align}
After separating the reciprocal space contributions that are longitudinal and transverse in the references frame of the particle, we can write 
\begin{align}\label{eqn:FWtK}
\PF & 
=\frac{1}{mV}\sum_{\mathbf{K}}l(\mathbf{K})\overline{W_l(\mathbf{K})}+t(\mathbf{K})\overline{W_t(\mathbf{K})},
\end{align}
where the overline denotes the complex conjugate. Here, we have implicitly defined the weights
\begin{align}
\label{eqn:WtK}
W_t(K)\equiv&\frac{\prefix}{(K\AAA)^{\frac{d}{2}-1}}\left[\phi J_{\frac{d}{2}-1}(K\AAA)+\frac{\psi}{K\AAA}J_{\frac{d}{2}}(K\AAA)\right]\\
=&\frac{\prefix}{(K\AAA)^{\frac{d}{2}-1}}\left[\left(\phi+\frac{\psi}{d}\right)J_{\frac{d}{2}-1}(K\AAA)+\frac{\psi}{d}J_{\frac{d}{2}+1}(K\AAA)\right],\nonumber\\
W_l(K)\equiv&\frac{\prefix}{(K\AAA)^{\frac{d}{2}-1}}\nonumber\\
&\times\left[\left(\phi+\frac{\psi}{d}\right)J_{\frac{d}{2}-1}(K\AAA)-(d-1)\frac{\psi}{d}J_{\frac{d}{2}+1}(K\AAA)\right].\nonumber
\end{align}
with   $\prefix=\frac{(2\pi)^{\frac d2}}{\rho G(\AAA)S_{d-1}}$, as well as the solvent momentum fluctuation vectors
\begin{align}
l(\mathbf{K})&\equiv\tvec{p}(\mathbf{K})\cdot(\mathbf{K}\otimes\mathbf{K})\label{eq:lk},\\
t(\mathbf{K})&\equiv\tvec{p}(\mathbf{K})\cdot(\mathbf{I}-\mathbf{K}\otimes\mathbf{K}).
\label{eq:tk}\end{align} 
Note that these vectors involve: (i) the positions of both the (stationary) 
Brownian particle and the solvent particles, and (ii) the momenta of the solvent particles. 
Interestingly, these vectors evolve with 
a projected dynamics, which -- after following the approximations outlined in this section -- amounts to that of a standard solvent in presence of a stationary Brownian particle. 

Using Eq.~\eqref{eqn:FWtK} we can then write the frequency domain version of the correlation function as
\begin{widetext}
\begin{align}\label{eqn:XXavg} 
	\scp{\PF(\omega)\cdot \PF}=
	\frac{1}{(mV)^2}\sum_{\mathbf{K}_1,\mathbf{K}_2} \vect{\overline{W_l(\vec{K}_1)} & \overline{W_t(\vec{K}_1)}} 
	\vect{\scp{l(\vec{K}_1,\omega)\cdot l(\vec{K}_2)} &  \scp{l(\vec{K}_1,\omega)\cdot t(\vec{K}_2)} \\
		  \scp{t(\vec{K}_1,\omega)\cdot l(\vec{K}_2)} & \scp{t(\vec{K}_1,\omega)\cdot t(\vec{K}_2)}}  \vect{{W_l(\vec{K}_2)}\\ {W_t(\vec{K}_2)}}.
\end{align} 
\end{widetext}
We have thus expressed the friction coefficient experienced by the 
Brownian particle (given in Eq.~\eqref{friction}) in terms of the statistical properties of the fluctuations of the
solvents momentum (relative to the stationary Brownian particle).

\section{Solvent Dynamics}
\label{sec:solvent}
In this section we consider the dynamics of the solvent momentum fluctuations. 
Our treatment uses the standard projection operator analysis of fluid dynamics, but is complicated by the presence of a stationary Brownian particle in its midst.

\subsection{Projection operator analysis of the solvent dynamics}

The solvent variables already present in Eq.~\eqref{eqn:XXavg} are the fluctuations of the 
longitudinal and transverse momenta. In the conventional procedure (e.g., to derive the equations 
of linearized hydrodynamics for a fluid\cite{Kadanoff:1963,Hansen:1986}) one would
also consider the number density fluctuations. Here, we follow Ref.~\onlinecite{Masters:1981} and consider instead the 
the spectrum of solvent density fluctuations relative to the particle position,
\begin{equation}
b(\mathbf{K})=\left[n(\mathbf{K})N(-\mathbf{K})-\langle n(\mathbf{K})N(-\mathbf{K})\rangle\right]\hat{\mathbf{K}}.
\end{equation}
Despite the similitude with the number density, $b(\vec{K})$ is a vector (not a scalar), and is thus not as straightforward to handle. 
If it's any consolation, this choice at least simplifies the notation, because then all three fluctuating quantities, $b$, $l$ and $t$, are vectors. 
Note that this treatment neglects potential couplings to the energy, which is also locally conserved. 
As pointed out in Note [14] of Ref.~\onlinecite{Masters:1981} assuming an incompressible fluid (as we do below) results in the energy not affecting the friction coefficient.

Like MM, we arrange these three variables, for all wavevectors of interest, as three vectors of vectors,
\begin{equation}
\vec{b}=\left(\begin{array}{c}
b(\vec{K}_1) \\ 
b(\vec{K}_2) \\ 
b(\vec{K}_3) \\ 
\vdots
\end{array}\right),\;\vec{l}=\left(\begin{array}{c}
l(\vec{K}_1) \\ 
l(\vec{K}_2) \\ 
l(\vec{K}_3) \\ 
\vdots
\end{array}\right),\;\vec{t}=\left(\begin{array}{c}
t(\vec{K}_1) \\ 
t(\vec{K}_2) \\ 
t(\vec{K}_3) \\ 
\vdots
\end{array}\right),
\end{equation}
which we compactly denote $\vec{c}=(\vec{b},\vec{l},\vec{t})^\intercal$.
This treatment also differs from conventional approaches, which usually consider a single wavevector. This added complexity cannot be avoided, because the different wavevectors are coupled as a result of translational symmetry being broken by
the presence of the (still stationary) Brownian particle. Thus, in the following we analyze the complete correlation matrix $\scp{\vec{c}(t) \vec{c}^\dag}$. Note that (following MM) we use the convention that the element $(\vec{K}_1,\vec{K}_2)$ of the $\scp{\vec{l} \vec{l}^\dag}$ sector of the correlation matrix is equal to average of the scalar product of two vectors, $\scp{l(\vec{K}_1)\cdot l(\vec{K}_2)}$.

To evaluate the correlation matrix $\scp{\vec{c}(t) \vec{c}^\dag}$,
we use the projection operator technique once again. This time the set of slow 
variables is the set of vectors $\vec{c}$.  Recall that the time (or frequency) dependence of this
correlation matrix is given by the projected dynamics given in Eq.~\eqref{projecttimederF}. Recall also that after the
approximations made in Sec.~\ref{sec:particle}, this projected dynamics amounts to the standard dynamics of the solvent in presence of a stationary particle. 

The projection operator expression for $\scp{\vec{c}(t) \vec{c}^\dag}$ (or, more conveniently, for its frequency domain counterpart, $\scp{\vec{c}(\omega) \vec{c}^\dag}$) is obtained as in Sec.~9.1 of Ref.~\onlinecite{Hansen:1986} 
\begin{equation}\label{eqn:ccomega}
        \scp{\vec{c}(\omega) \vec{c}^\dag}=\scp{\vec{c} \vec{c}^\dag}\RR(\omega)\scp{\vec{c} \vec{c}^\dag},
\end{equation}
where the memory matrix $\RR$ is given by 
\begin{equation}\label{eqn:RRomega}
        \RR(\omega)=\Bigl(-i\omega \scp{\vec{c} \vec{c}^\dag}-\scp{\dot{\vec{c}} \vec{c}^\dag}
+\scpb{\dot{\vec{c}}(\omega), \dot{\vec{c}}^\dag}\Bigr)^{-1}.
\end{equation}
Except for the normalization convention, this result is thus equivalent to Eq.~(9.1.42) in Ref.~\onlinecite{Hansen:1986}. 
Note that for compactness, the notation in the above two equations follows that of MM, even though it is potentially confusing. In particular, the dot in the 
static expressions, $\scp{\dot{\vec{c}} \vec{c}^\dag}$, denotes the standard time derivative, while the dot in the 
dynamical expressions, $\scpb{\dot{\vec{c}}(\omega), \dot{\vec{c}}^\dag}$, denotes the projected time derivative. It follows from manipulations of the projection operator that the time (or frequency) dependence in 
$\scpb{\dot{\vec{c}}(\omega), \dot{\vec{c}}^\dag}$ is then given by the projected dynamics. That is, for arbitrary vectors $\vec{u}$ and $\vec{v}$, 
\begin{equation}
    \scpb{\dot{\vec{u}},\dot{\vec{v}}^\dag}(t)=\scp{e^{i\QQ_c\LL t}(\QQ_c\dot{\vec{u}}) \QQ_c \dot{\vec{v}}^\dag},
\end{equation}
and
\begin{equation}
    \scpb{\dot{\vec{u}}(\omega),\dot{\vec{v}}^\dag}=\int_0^\infty \dd t\ e^{i\omega t}[\dot{\vec{u}},\dot{\vec{v}}^\dag](t),
\end{equation}
where $\QQ_c$ denotes the projection operator on the space orthogonal to $\vec{c}$ and the Liouville operator describes the standard solvent dynamics in presence of the immobile (Brownian) particle, as approximated in Sec.~\ref{sec:particle}. 
Since $\scp{\vec{l} \vec{b}^\dag}=0$ and $\scp{\vec{t} \vec{b}^\dag}=0$, 
computing the correlation function in Eq.~\eqref{eqn:XXavg} only requires evaluating
\begin{widetext}
\begin{equation}\label{eqn:restricted-ccomega}
	\vect{ \scp{\vec{l}(\omega) \vec{l}^\dag} & \scp{\vec{l}(\omega) \vec{t}^\dag}\\
		    \scp{\vec{t}(\omega) \vec{l}^\dag} &\scp{\vec{t}(\omega) \vec{t}^\dag}} = 
	\vect{ \scp{\vec{l} \vec{l}^\dag} & \scp{\vec{l} \vec{t}^\dag}\\
		    \scp{\vec{t} \vec{l}^\dag} &\scp{\vec{t} \vec{t}^\dag}}
	\vect{\RR_{ll}(\omega)& \RR_{lt}(\omega)\\ \RR_{tl}(\omega)& \RR_{tt}(\omega)}
	\vect{ \scp{\vec{l} \vec{l}^\dag} & \scp{\vec{l} \vec{t}^\dag}\\
		    \scp{\vec{t} \vec{l}^\dag} &\scp{\vec{t} \vec{t}^\dag}},
\end{equation}
so, for example, the element $(\vec{K}_1,\vec{K}_2)$ of $\scp{\vec{l}(\omega) \vec{l}^\dag}$ has the form
\begin{align*}
\scp{l(\vec{K}_1,\omega)\cdot l(\vec{K}_2)}&=\sum_{\vec{K}',\vec{K}''} \scp{l(\vec{K}_1)\cdot l(\vec{K}')}\RR_{ll}(\vec{K}',\vec{K}'',\omega)\scp{l(\vec{K}'')\cdot l(\vec{K}_2)}
+\scp{l(\vec{K}_1)\cdot t(\vec{K}')}\RR_{tl}(\vec{K}',\vec{K}'',\omega)\scp{l(\vec{K}'')\cdot l(\vec{K}_2)}\notag\\
&\phantom{=\sum}\scp{l(\vec{K}_1)\cdot l(\vec{K}')}\RR_{lt}(\vec{K}',\vec{K}'',\omega)\scp{t(\vec{K}'')\cdot l(\vec{K}_2)}+\scp{l(\vec{K}_1)\cdot t(\vec{K}')}\RR_{tt}(\vec{K}',\vec{K}'',\omega)\scp{t(\vec{K}'')\cdot l(\vec{K}_2)}.
\end{align*}
After gathering these terms, it is fortunately possible to simplify some of the resulting integrals by 
using the real-space transformation described in Appendix~\ref{sec:MMAppB}.
The combination of Eqs.~\eqref{eqn:XXavg} and 
\eqref{eqn:restricted-ccomega} then becomes 
\begin{align}\label{eqn:XXfinal}
\langle \PF(\omega)\cdot\PF\rangle=\left[\frac{\rho G(\AAA)V}{\beta(2\pi)^d}\right]^2\int \dd\vec{K}_1&\dd\vec{K}_2
[(d-1)^2 W_t(K_1)W_t(K_2)\RR_{tt}(\vec{K}_1,\vec{K}_2,\omega)\notag
+(d-1)W_t(K_1)W_l(K_2)\RR_{tl}(\vec{K}_1,\vec{K}_2,\omega)\\
&+(d-1)W_l(K_1)W_t(K_2)\RR_{lt}(\vec{K}_1,\vec{K}_2,\omega)+W_l(K_1)W_l(K_2)\RR_{ll}(\vec{K}_1,\vec{K}_2,\omega)].
\end{align}
The remaining task of obtaining the various elements of $\RR^{-1}$ is described in the next two subsections.
\end{widetext}
\subsection{Static averages} \label{sec:staticavg}

We first evaluate the various static averages of $\vec{c}$ and $\dot{\vec{c}}$. Using the structural quantities defined in Sec.~\ref{sec:static} we can show that all terms in matrices $\langle\vec{c}\vec{c}^\dag\rangle$ and $\langle\dot{\vec{c}} \vec{c}^\dag\rangle$ vanish, except for
\begin{align*}
\langle l(\vec{K})\cdot l(\vec{K}')\rangle&=\frac{m}{\beta}(\hat{\vec{K}}\cdot\hat{\vec{K}}')^2[S(\vec{K}-\vec{K}')-1+N_\mathrm{f}\delta_{\vec{K},\vec{K}'}]\\
\langle \dot{b}(\vec{K})\cdot l(\vec{K}')\rangle&=\frac{iK}{m}\langle l(\vec{K})\cdot l(\vec{K}')\rangle\\
\langle l(\vec{K})\cdot t(\vec{K}')\rangle&=\frac{m}{\beta}[1-(\hat{\vec{K}}\cdot\hat{\vec{K}}')^2][S(\vec{K}-\vec{K}')-1]\\
\langle \dot{b}(\vec{K})\cdot t(\vec{K}')\rangle&=\frac{iK}{m}\langle l(\vec{K})\cdot t(\vec{K}')\rangle\\
\langle t(\vec{K})\cdot t(\vec{K}')\rangle&=\frac{m}{\beta}[(d-2)+(\hat{\vec{K}}\cdot\hat{\vec{K}}')^2]\\
&\times[S(\vec{K}-\vec{K}')-1+N_\mathrm{f}\delta_{\vec{K},\vec{K}'}]\\
\langle b(\vec{K})\cdot b(\vec{K}')\rangle&=(\hat{\vec{K}}\cdot\hat{\vec{K}}')
\left\{\left\langle\sum_{i j}e^{-i(\vec{K}-\vec{K}')\cdot\vec{R}}e^{i\vec{K}\cdot\vec{r}_j-i\vec{K}'\cdot\vec{r}_i}\right\rangle\right.\\
&-
\left.[S(\vec{K})-1+N_\mathrm{f}\delta_{\vec{K},\vec{0}}][S(\vec{K}')-1+N_\mathrm{f}\delta_{\vec{K}',\vec{0}}]\right\}.
\end{align*}
Note that in this last expression off-diagonal terms involve three-body correlations that remain of order unity, while diagonal terms are extensive. In the thermodynamic limit, the solvent density correlator thus simplifies to its bulk expression
\begin{align}
\langle b(\vec{K})\cdot b(\vec{K}')\rangle= \delta_{\vec{K},\vec{K}'} N_\mathrm{f} s(\vec{K}). 
\end{align}

\subsection{Dynamical averages}
We next consider the
elements of the dynamical matrix $\scpb{\dot{\vec{c}}(\omega), \dot{\vec{c}}^\dag}$. Recall that the vectors $l(\vec{K})$ and $t(\vec{K})$ are
the longitudinal and transverse components, relative to the Brownian particle, of the solvent momentum, 
$\tilde{\vec{p}}(\vec{K})\equiv\vec{p}(\vec{K})N(-\vec{K})$ (see Eqs.~\eqref{eq:lk} and \eqref{eq:tk}). 
Neglecting the time dependence of the particle position (as motivated in Sec.~\ref{sec:particle}) and the effect of the particle on the fluid, we get
\begin{equation}
\dot{\tilde{\vec{p}}}(\vec{K}) = i \vec{\sigma}(\vec{K})\cdot \vec{K} N(-\vec{K}),
\end{equation}
where the microscopic stress tensor 
\begin{align*}\label{eqn:microscopicstresstensor}
        \vec{\sigma}(\vec{K}) & = \sum_i e^{i\vec{K}\cdot \vec{r}_i} 
\Bigl(\frac{\vec{p}_i \otimes \vec{p}_i}{m}
        +\frac12\sum_j\frac{1-e^{i\vec{K}\cdot \vec{r}_{ij}}}{i\vec{K}\cdot \vec{r}_{ij}}\vec{f}_{ij}\otimes\vec{r}_{ij}\Bigr),
\end{align*}
includes a contribution that arises from the microscopic force between solvent particles, 
$\vec{f}_{ij}=-\partial_{\vec{r}_i} u(|\vec{r}_i-\vec{r}_j|)$.

Recall also that the elements of the memory matrix, $\RR$, are expressed in terms of time derivatives projected 
on the subspace orthogonal to $\vec{c}$. 
Because correlations between the components of the solvent momentum and the stress tensor vanish by symmetry, 
the projected time derivatives of $l(\vec{K})$ and $t(\vec{K})$ involve the projected stress tensor 
$\vec{\sigma}^c(\vec{K})$,
\begin{align}
\vec{\sigma}^c(\vec{K}) =
\vec{\sigma}(\vec{K})-\frac{n(\vec{K})}{\beta\;s(\vec{K})}\vec{I}
\equiv \sum_i e^{i\vec{K}\cdot \vec{r}_i}\vec{\zeta}_i(\vec{K}),
\end{align}
where last equality implicitly defines single-particle contributions to the stress tensor, 
$\vec{\zeta}_i$. 

We are now about to formulate the most important approximation of this work, taking a slightly different approach than MM. We illustrate the process by considering the component $\left\langle\dot{t}(\vec{K},\omega=0)\cdot \dot{t}(\vec{K}')\right\rangle$ of the full matrix $\scpb{\dot{\vec{c}}(\omega=0), \dot{\vec{c}}^\dag}$:
\begin{widetext}
\begin{align}\label{eqn:tt}
& \left\langle\dot{t}(\vec{K},\omega=0)\cdot \dot{t}(\vec{K}')\right\rangle \\ \nonumber &=
\int_0^\infty \dd\tau \left\langle \sum_{i,j} e^{i\vec{K}\cdot (\vec{r}_i(\tau)-\vec{R}(\tau))}
e^{-i\vec{K}'\cdot (\vec{r}_j(0)-\vec{R}(0))}
\left[(\vec{I}-(\hat{\vec{K}}\otimes\hat{\vec{K}})) \cdot
\vec{\zeta}_i(\vec{K};\tau)\cdot \vec{K} \right]\cdot
\left[(\vec{I}-(\hat{\vec{K}}'\otimes\hat{\vec{K}}')) \cdot \vec{\zeta}_j(-\vec{K}';0)
\cdot \vec{K}' \right]\right\rangle,
\end{align}
where $\tau$ denotes the projected time evolution. 
Equation~\eqref{eqn:tt} contains two elements. The first, 
$\sum_{i,j} e^{i\vec{K}\cdot (\vec{r}_i(\tau)-\vec{R}(\tau))}
e^{-i\vec{K}'\cdot (\vec{r}_j(0)-\vec{R}(0))}$, describes the evolution of the solvent density around the 
Brownian particle. The second, which involves $\vec{\zeta}_i(\vec{K};\tau)\vec{\zeta}_j(-\vec{K}';0)$, 
describes the evolution of the solvent stress tensor fluctuations. 

We assume that the time and length scales of these two elements can be separated. Specifically, we assume that the integrand in expression $\left\langle\dot{t}(\vec{K},\omega=0)\cdot \dot{t}(\vec{K}')\right\rangle$ can be factorized as
\begin{align}\label{eqn:ttfact}
& 
\left\langle \sum_{i,j} e^{i\vec{K}\cdot (\vec{r}_i(\tau)-\vec{R}(\tau))}
e^{-i\vec{K}'\cdot (\vec{r}_j(0)-\vec{R}(0))}
\left[(\vec{I}-(\hat{\vec{K}}\otimes\hat{\vec{K}})) \cdot
\vec{\zeta}_i(\vec{K};\tau)\cdot \vec{K} \right]\cdot
\left[(\vec{I}-(\hat{\vec{K}}'\otimes\hat{\vec{K}}')) \cdot \vec{\zeta}_j(-\vec{K}';0)
\cdot \vec{K}' \right]\right\rangle \\ & \nonumber \approx
N_\mathrm{f}^{-1} \left\langle \sum_{i,j} e^{i\vec{K}\cdot (\vec{r}_i(0)-\vec{R}(0)}
e^{-i\vec{K}'\cdot (\vec{r}_j(0)-\vec{R}(0))}\right\rangle \left\langle \sum_{i,j} \left[(\vec{I}-(\hat{\vec{K}}\otimes\hat{\vec{K}})) \cdot
\vec{\zeta}_i(\vec{0};t)\cdot \vec{K} \right]\cdot
\left[(\vec{I}-(\hat{\vec{K}}'\otimes\hat{\vec{K}}')) \cdot \vec{\zeta}_j(\vec{0};0)
\cdot \vec{K}' \right]\right\rangle.
\end{align}
Equation~\eqref{eqn:ttfact} combines several approximations that are made concurrently. First, we assume that the relaxation of stress fluctuations is not affected by the presence of the Brownian particle. The configurational averages can thus be factorized. Second, we assume that solvent density around the particle evolves slowly compared to the stress fluctuations. The solvent density correlations can thus be evaluated concurrently, i.e.,
\begin{equation}\label{eqn:nNnN}
\left\langle \sum_{i,j} e^{i\vec{K}\cdot (\vec{r}_i(\tau)-\vec{R}(\tau))}
e^{-i\vec{K}'\cdot (\vec{r}_j(0)-\vec{R}(0))}\right\rangle \approx
\left\langle \sum_{i,j} e^{i\vec{K}\cdot (\vec{r}_i(0)-\vec{R}(0))}
e^{-i\vec{K}'\cdot (\vec{r}_j(0)-\vec{R}(0))}\right\rangle
\approx N_\mathrm{f} s(\vec{K})\delta_{\vec{K}\vec{K}'} + S(\vec{K}-\vec{K}')-1.
\end{equation}
Note that this approximation is slightly different from MM's, in that it preserves information about the solvent structure.
Third, we assume that the length scale for solvent and stress correlations are well separated. In other words, we assume that the wavevectors at which the correlations of the solvent around the Brownian particle are non-trivial are small compared to those relevant to the stress correlations. It is thus possible to take the $\vec{K}\to\vec{0}$ limit of $\vec{\zeta}_i(\vec{K};\tau)$, and in that limit the projected time evolution of the stress correlations can be replaced by their standard time evolution,\cite{Ernst:1975} i.e., $\tau\rightarrow t$.

Integrating the stress tensor correlations at zero wavevector is achieved with the help of the following relations for the elements of the rank 2 viscosity tensor,
\begin{equation}
\bm{\eta}(\vec{0},\omega=0)=\lim_{\vec{K}\to\vec{0}} \frac{\beta}{V} \int_0^\infty\dd t 
\langle\vec{\sigma}^c(\vec{K},t)\otimes\vec{\sigma}^c(-\vec{K})\rangle,
\end{equation}
in terms of the bulk and shear viscosity, $\eta_\mathrm{B}$ and $\eta_\mathrm{S}$, respectively (see Appendix~\ref{sec:etatensor}), 
\begin{equation}\label{eqn:visc}\begin{aligned}
\eta_{iiii}(\vec{0},\omega=0)&=\eta_\mathrm{B}+\frac{2(d-1)}{d}\eta_\mathrm{S},\\
\eta_{ijij}(\vec{0},\omega=0)&=\eta_\mathrm{S}\;\;\;\mathrm{for}\;i\neq j,\\
\eta_{iijj}(\vec{0},\omega=0)&=\eta_\mathrm{B}-\frac{2}{d}\eta_\mathrm{S} \;\;\;\mathrm{for}\;i\neq j.
\end{aligned}\end{equation}
Using Eqs.~\eqref{eqn:visc} gives for the time integral of the second configurational average in Eq.~\eqref{eqn:ttfact},
\begin{align}\label{eqn:ssint}
 N_\mathrm{f}^{-1} \int_0^\infty \dd t \left\langle \sum_{i,j} \left[(\vec{I}-(\hat{\vec{K}}\otimes\hat{\vec{K}})) \cdot
\vec{\zeta}_i(\vec{0};t)\cdot \vec{K} \right]\cdot
\left[(\vec{I}-(\hat{\vec{K}}'\otimes\hat{\vec{K}}')) \cdot \vec{\zeta}_j(\vec{0};0)
\cdot \vec{K}' \right]\right\rangle
= \frac{(\vec{K}\cdot \vec{K}')}{\beta\rho}
\left[d-3+2(\hat{\vec{K}}\cdot \hat{\vec{K}}')^2\right]\eta_\mathrm{S}.
\end{align}

Combining Eqs.~\eqref{eqn:nNnN} and \eqref{eqn:ssint}, we finally obtain the approximate expression 
\begin{align}
\left\langle\dot{t}(\vec{K},\omega=0)\cdot \dot{t}(\vec{K}')\right\rangle & \approx 
\left[N_\mathrm{f} s(\vec{K})\delta_{\vec{K}\vec{K}'} + S(\vec{K}-\vec{K}')-1 \right] 
\frac{(\vec{K}\cdot \vec{K}')}{\beta\rho}
\left[d-3+2(\hat{\vec{K}}\cdot \hat{\vec{K}}')^2\right]\eta_\mathrm{S}.
\end{align}
Repeating the above procedure for the other elements gives
\begin{align}
\left\langle\dot{l}(\vec{K},\omega=0)\cdot\dot{l}(\vec{K}')\right\rangle & \approx
\left[N_\mathrm{f} s(\vec{K})\delta_{\vec{K}\vec{K}'} + S(\vec{K}-\vec{K}')-1 \right] 
\frac{(\vec{K}\cdot\vec{K}')}{\beta \rho}
\left\{\eta_\mathrm{B}+\left[2(\hat{\vec{K}}\cdot\hat{\vec{K}}')^2-\frac2d\right]\eta_\mathrm{S}\right\}
\\
\left\langle\dot{l}(\vec{K},\omega=0)\cdot \dot{t}(\vec{K}')\right\rangle &=
\left\langle\dot{t}(\vec{K},\omega=0)\cdot \dot{l}(\vec{K}')\right\rangle \approx
\left[S(\vec{K}-\vec{K}')-1 \right] \frac{2(\vec{K}\cdot \vec{K}')}{\beta\rho}
\left[1-(\hat{\vec{K}}\cdot \hat{\vec{K}}')^2\right]\eta_\mathrm{S}.
\end{align}
\end{widetext}
Note that the above four expressions preserve off-diagonal terms, \textit{i.e.} with $\vec{K}\neq\vec{K}'$. These terms capture the breaking of the translational symmetry of the solvent caused by the sole presence of the Brownian particle. As we will see below, keeping these terms is essential for reproducing the SER. 

The only remaining task consists of performing the inversion given in Eq.~\eqref{eqn:RRomega} to obtain $\RR$ and solve Eq.~\eqref{eqn:XXfinal}. This challenging program
is achieved for a couple of solvation models in the following two sections.

\section{Unphysical model}
\label{sec:unphysical}
The simplest possible model of solvation is to pose that the solvent can penetrate the Brownian particle, i.e., $G(r)=1$ and $S(\vec{K})-1=0$, and that the solvent has the structure of an ideal gas, i.e., $g(r)=\Theta(r-2a)$ and
\begin{align}
\label{eqn:solvstruct}
s(K)&=1-\bar\varphi\Gamma\bigl(\frac d2+1\bigr)\frac{J_{\frac d2}(2Ka)}{(Ka)^{\frac d2}}\\
&=1-\bar{\varphi}\{1+\mathcal{O}[(Ka)^2/d]\},\nonumber
\end{align}
where $\bar{\varphi}=2^d\varphi$ is the natural scale for the solvent volume fraction, $\varphi=\rho V_{d}(a)$, for particles of volume $V_d(a)$~\cite{Charbonneau:2013}. For this solvation model the only non-vanishing correlators are 
\begin{align*}
&\scp{b(\vec{K})\cdot b(\vec{K}')}=N_\mathrm{f}\delta_{\vec{K},\vec{K}'}\\
&\scp{l(\vec{K})\cdot l(\vec{K}')}=\frac{mN_\mathrm{f}}{\beta}\delta_{\vec{K},\vec{K}'}\\
&\scp{t(\vec{K})\cdot t(\vec{K}')}=\frac{m(d-1)N_\mathrm{f}}{\beta}\delta_{\vec{K},\vec{K}'}\\
&\scp{\dot{b}(\vec{K})\cdot l(\vec{K}')}=\frac{i KN_\mathrm{f}}{\beta}\delta_{\vec{K},\vec{K}'}\\
&\scp{\dot{l}(\vec{K},\omega=0)\cdot \dot{l}(\vec{K}')}=\frac{N_\mathrm{f}s(K)K^2}{\beta\rho}\left[\eta_\mathrm{B}+2\frac{(d-1)}d\eta_\mathrm{S}\right]\delta_{\vec{K},\vec{K}'}\\
&\scp{\dot{t}(\vec{K},\omega=0)\cdot \dot{t}(\vec{K}')}= \frac{N_\mathrm{f}s(K)K^2}{\beta\rho}(d-1)\eta_\mathrm{S}\delta_{\vec{K},\vec{K}'}.
\end{align*}
Note that for this solvation model all the correlators are diagonal in $\vec{K}$. Translational symmetry is thus recovered.

Defining the diagonal matrices $\vec{D}_1=\diag(\ldots,K,\ldots)$, $\vec{D}_2=\diag(\ldots,s(K),\ldots)$, and $\vec{D}_3=\diag(\ldots,s(K) K^2,\ldots)$, we can thus write the matrix
\begin{widetext}
\begin{align}
\RR(\omega)^{-1}&= \vect{ -i\omega N_\mathrm{f}\vec{D}_2  &-\frac{iN_\mathrm{f}}{\beta}\vec{D}_1  &	\\
					-\frac{iN_\mathrm{f}}{\beta}\vec{D}_1& -\frac{imN_\mathrm{f}\omega}{\beta}\vec{I}&\\
					&&	-i\omega\frac{m(d-1)N_\mathrm{f}}{\beta}\vec{I}}+\vect{\vec{0}&& \\
								&\frac{N_\mathrm{f}}{\beta\rho}(\eta_\mathrm{B}+2\frac{(d-1)}d\eta_\mathrm{S})\vec{D}_3&\\
								&& \frac{N_\mathrm{f}(d-1)\eta_\mathrm{S}}{\rho\beta}\vec{D}_3}.
\end{align}
Given that $\vec{D}_1,\vec{D}_2,\vec{D}_3$ commute, one can treat $\RR(\omega)^{-1}$ as a $3\times 3$ matrix and straightforwardly compute $\RR(\omega)$. Substituting the transverse and longitudinal sectors of $\RR(\omega)$ into Eq.~\eqref{eqn:XXfinal} we then get
\begin{align}
\langle \PF(\omega)\cdot\PF\rangle=
\frac{G(\AAA)^2 \rho S_{d-1}}{\beta (2\pi)^d}\int_0^\infty\dd K K^{d-1}\left[\frac{(d-1) W_t^2(K)}{-i\omega m+\frac{\eta_\mathrm{S} s(K)K^2}{\rho}}+\frac{W_l^2(K)}{-i\omega m+\frac{iK^2}{\omega\beta}+\frac{(\eta_\mathrm{B}+2\frac{d-1}{d}\eta_\mathrm{S})s(K) K^2}{\rho}}\right].
\end{align}
Using once more the incompressible solvent assumption, i.e., $\nabla \vec{p}=0$ and thus $\vec{K}\cdot\vec{p}(\vec{K})=0$ for all $\vec{K}$, implies that $W_l(K)=0$. For the slip case, we then obtain (using Eq.~\eqref{eqn:int J2 over z z2-x2})
\begin{align*}
\langle\PF(\omega)\cdot\PF\rangle&=\frac{\psi^2(d-1)}{\beta\eta_\mathrm{S} S_{d-1}\AAA^{d-2}}\int_0^\infty \dd K K^{-1} \frac{J_{\frac{d}{2}}^2(K)}{-\delta^2+K^2\Bigl(1-\bar{\varphi}\Gamma\left(\frac{d}{2}+1\right)\frac{J_{\frac{d}{2}}(2Ka/\AAA)}{(Ka/\AAA)^{\frac{d}{2}}}\Bigr)}\notag\\
&=\frac{\psi^2(d-1)}{\beta d\eta_\mathrm{S} S_{d-1}\AAA^{d-2}}\frac{i\pi}{2\delta}\Bigl[H^{(1)}_{\frac d2-1}(\delta)J_{\frac d2}(\delta)+H^{(1)}_{\frac d2}(\delta)J_{\frac d2+1}(\delta)\Bigr][1+\mathcal{O}(\bar{\varphi})],
\end{align*}
\end{widetext}
where $\delta=\sqrt{i\omega m\rho\AAA^2/\eta_\mathrm{S}}$ is a materials constant for a fixed $\omega$.
Plugging this result into Eqs.~\eqref{eq:mxiboundary} and considering a low-density solvent, $\bar{\varphi}\ll 1$, in the zero-frequency limit gives
\begin{equation}
M\xi_\mathrm{unphys}(0;0,\psi)=\frac{(d^2-4)S_{d-1}\eta_\mathrm{S} \AAA^{d-2}}{2(d-1)}.
\end{equation}

Remarkably, this result coincides with that of Keyes and Oppenheim~\cite{Keyes:1973} for $d=3$, i.e., $M\xi'(0;0)=5\pi\AAA\eta_\mathrm{S}$, and with the Hadamard--Rybczynski drag on a sphere of fluid of viscosity $\eta_\mathrm{S}$ by a fluid of the same viscosity (see Appendix~\ref{sec:drag_on_a_drop_of_fluid}). It thus extends to all dimensions the conjecture made by MM for this case in $d=3$. In order to obtain a firmer grasp on the validity of this analogy the finite-frequency scaling of this problem should also be considered. The corresponding hydrodynamic calculation has, however, not yet been attempted. We thus leave this particular question to a later study and proceed instead with a different solvation model.

\section{Physical model}
\label{sec:physical}
A more physical description of solvation should at least enforce volume exclusion by the Brownian particle. As a minimal model, we thus take $G(r)=\Theta(r-\AAA)$ and $g(r)=\Theta(r-2a)$. Even if we assume an incompressible solvent from the start, this choice is highly nontrivial to analyze. The full solution of
\begin{widetext}
\begin{equation}
\langle\PF(\omega)\cdot\PF\rangle=\left[\frac{(d-1)\rho G(\AAA)V}{\beta (2\pi)^d}\right]^2\iint\dd \vec{K}_1\dd \vec{K}_2W_t(\vec{K}_1)\mathcal{R}(\vec{K}_1,\vec{K}_2,\omega)W_t(\vec{K}_2),
\end{equation}
and its derivation are presented in Appendix~\ref{sec:MMAppC}. While obtaining $\RR$ from its inverse $\RRi$ is straightforward for the unphysical model, here the problem is not similarly tractable. In order to make progress, the key insight is that one doesn't need the full matrix $\RR$ to compute the self-correlation $\scp{\PF(\omega)\cdot\PF}$. One only needs the partial inverse
$\tilde\wp(\vec{K}\AAA)\equiv \int \!\!\dd\vec{K}'\RR(\vec{K},\vec{K}')W_t(K')$, and in fact only the two coefficients
\begin{equation*}
	B_\pm\equiv \frac1{S_{d-1}}\int \!\!\!\dd\vec{K}\frac{J_{\frac d2\pm1}(K\AAA)}{(K\AAA)^{\frac d2-1}}\tilde\wp(\vec{K}\AAA).
\end{equation*}
Schematically, because $\tilde\wp = \RR W$, we have $W=\RRi \tilde\wp$. In order to make use of Hankel's theorem,
$\int_0^\infty \dd rrJ_\nu(Kr)J_\nu(K'r)=\frac{\delta(K-K')}{K}$, we temporarily define $G_\epsilon(r)\equiv 1-\epsilon\Theta(\AAA-r)$, which produces two parts in the integral $\int\!\!\RRi(\vec{K},\vec{K}')\tilde\wp(\vec{K}'\AAA)$: one over all space, and one on the ball of radius $\AAA$. In other words, we schematically have
\begin{equation}
	W(K) \propto\bigl((K\AAA)^2-\delta^2\bigr)\tilde\wp(\vec{K}\AAA)-\epsilon \Upsilon_\AAA.
\end{equation}
The resulting identity
\begin{equation}
	\tilde\wp(\vec{K}\AAA) \propto \frac{W(K)}{(K\AAA)^2-\delta^2} +\epsilon \frac{\Upsilon_\AAA}{(K\AAA)^2-\delta^2}
\end{equation}
 is  recursive because the integrand in $\Upsilon_\AAA$ contains $\tilde\wp$, which at first may seem intractable. Yet, integrating once more against a proper multiple of $J_{\frac d2\pm 1}$ and using various Bessel identities, two linear equations for $B_\pm$ can be obtained. One can thus solves this system. Finally setting $\epsilon=1$ gives the desired final expression.

In the zero-frequency limit this analysis further simplifies to
\begin{equation}
\langle\PF(0)\cdot\PF\rangle=\frac{(d-1)}{\beta\eta_\mathrm{S} S_{d-1}\AAA^{d-2}}\left[\frac{1}{(d-2)}\left(\phi+\frac{\psi}{d}\right)^2+\frac{1}{2}\left(\frac{\psi}{d}\right)^2\right]\left[1+\bar{\varphi}\bigl(1+\mathcal{O}(a/\AAA)\bigr)\right].
\end{equation}
\end{widetext}
In the low density limit, we then obtain [from Eq.~\eqref{eq:mxiboundary}] the friction coefficients
\begin{align}
\label{eqn:stickslip}
M\xi_\mathrm{phys}(0;0,\psi)&=\frac{8\pi^{d/2}\eta_\mathrm{S}\AAA^{d-2}}{(d-1)\Gamma(d/2-1)},\\
M\xi_\mathrm{phys}(0;\phi;0)&=\frac{4d\pi^{d/2}\eta_\mathrm{S}\AAA^{d-2}}{(d-1)\Gamma(d/2-1)},
\end{align}
for the slip and stick boundary conditions, respectively. Remarkably, these results precisely correspond to those obtained by generalizing the standard hydrodynamic analysis to arbitrary $d$ given in Eqs.~(44) and (48) of Ref.~\onlinecite{Charbonneau:2013}. By preserving the off-diagonal contributions, the improved solvent model thus robustly captures the essential physics behind the SER. 

Interestingly, it also correctly predicts that the case $d=2$ is singular, in that the friction coefficient vanishes. This is the reflection of the Stokes paradox known from macroscopic thermodynamics. Indeed, the projection operator formalism we used leads to a linearized hydrodynamics, and thus corresponds to a low Reynolds number limit. We note that a consistent treatment of the $d=2$ case requires also dealing with $t^{-1}$ long-time tails in the stress autocorrelation function \cite{Hansen:1986}. This analysis is beyond the scope of the present work. 

A couple of additional microscopic insights also emerge. First, it unambiguously identifies the hydrodynamic radius of the Brownian particle. Its proper asymptotic form is the sum of the particle radii, $\AAA=A+a$, which is where particles exchange momentum. This result thus confirms the proposal of Refs.~\onlinecite{Schmidt:2003,Charbonneau:2013}. Second, it identifies how the solvent structure qualitatively affects the SER. Systematic deviations from the hydrodynamic prediction are indeed expected for $\bar{\varphi}\gtrsim1$, because then friction on the Brownian particle diminishes significantly.
Although such contribution might be challenging to measure in experiments because of the difficulty of precisely determining $\AAA$, numerical simulations of hard sphere fluids suffer no such limitation. The extent to which finite-dimensional corrections to  $S(K)$ and other approximations might hide this effect, however, have yet to be determined. Solving for an even more realistic model of the solvent structure is thus a natural next step.

A more controlled limit is that of $d\rightarrow\infty$, in which the approximations done in Secs.~\ref{sec:particle} and \ref{sec:solvent} as well as the current approximations to $S(K)$ and $s(K)$ become exact. In that context, it is thus reasonable to extend the analysis to the limit of self-solvation, in which the Brownian particle radius is the same as that of the solvent particles, i.e., $A\rightarrow a$ and $\AAA\rightarrow\mathcal{D}\equiv2A$. For $\bar{\varphi}\gg1$, the slip case then reduces to 
\begin{align}
\label{eq:highdxi}
M\xi_\mathrm{phys}(0;0,\psi)&\approx\frac{2d\times 2^d (d - 2)  V_d(\mathcal{D})\eta_\mathrm{S}}{(d-1)\mathcal{D}^2\bar{\varphi}}\nonumber\\
&\overset{d\rightarrow\infty}{\cong} \frac{2\times 2^d V_d(\mathcal{D})\eta_\mathrm{S}}{\mathcal{D}^2 \widehat{\varphi}}.
\end{align}
where $\widehat{\varphi}=\bar{\varphi}/d$. Remarkably, Eq.~\eqref{eq:highdxi} recovers the nontrivial SER-like result of Maimbourg et al. in $d\rightarrow\infty$ fluids~\cite{Maimbourg:2016,Charbonneau:2017}. In this limit, this scaling is expected to hold from the low-density fluid regime up to the dynamical transition, $\widehat{\varphi}_\dd=4.8$. The solvent structure therefore serves as counterweight to viscosity in this case. 

The sign and magnitude of the above effects are also consistent with the dominant deviation from SER observed in numerical simulations of self-solvation in $d=3$-$8$~\cite{Schmidt:2003,Charbonneau:2013}. A more careful consideration of the separate role of density and solvent structure in the regime where $\bar{\varphi}>1$ and $0<a/\AAA\leq1$ should, however, be undertaken to confirm the microscopic origin of these discrepancies.

\section{Conclusion}
\label{sec:conclusion}
By refining and generalizing the microscopic derivation of SER by Masters and Madden~\cite{Masters:1981}, we have obtained a number of microsopic insights that had previously largely gone unnoticed, even in alternate microscopic derivations~\cite{Schofield:1992,Itami:2015}: (i) the unambiguous definition of the hydrodynamic radius; (ii) the leading microscopic correction to SER; and (iii) an independent derivation of the SER-like for self-solvation in $d\rightarrow\infty$. These predictions should motivate further numerical work, in order to systematically identify how the various corrections to the hydrodynamic SER develop with particle radius, solvent density, and dimension.

Following MM, we have here worked in reciprocal space. This scheme offers many technical advantages in problems with unbroken translational symmetry, and is also convenient in the present problem, where a solution is sought in the reference frame of the Brownian particle. The resulting expressions of Section \ref{sec:static} for the length of longitudinal and transverse components of the Fourier transformed momentum vector, for instance, are strikingly simple.  It would likely be instructive to repeat our analysis in real space, thence following more closely the macroscopic analysis~\cite{Charbonneau:2013}. Even though a comparably straightforward expression for the solvent momentum field is then obtained (see, e.g., Eq.~(30) of Ref.~\onlinecite{Charbonneau:2013}), translating one approach into the other appears to be rather technically difficult. This problem is thus left for a future study.

A key assumption of the current derivation is that the solvent is memoryless. As the solvent density increases, however, its dynamics grows dramatically sluggish, and its memory lengthens accordingly. In experiments and simulations the emergence of such memory is accompanied with a breakdown of SER; the diffusivity of solvent-size particles in deeply supercooled liquids is markedly larger than what SER predicts. Even larger deviations are observed when the particle radius is much smaller than the solvent, with the random Lorentz gas serving as a limit case~\cite{Jin:2015}. 
The standard theoretical approach for including memory to the solvent flow is through a mode-coupling scheme~\cite{Gotze:2009}. As mentioned in the introduction, however, existing schemes fail to recover key aspects of the physics of viscous liquids. Surprisingly, such approaches all neglect off-diagonal terms. Seeing how crucial these terms are to recover SER, it would be interesting to formulate a mode-coupling theory of viscous liquids that includes them. Hopefully, such herculean effort could provide a dimensionally consistent theory of viscous liquids as well as additional insight into the associated breakdown of SER.

\section*{Supplementary Material}
See supplementary material for a companion Maple 2017 code to compute the hydrodynamic drag on a drop of fluid suspended in a fluid with a different viscosity, in all dimensions.

\begin{acknowledgments}
We thank David Reichman and Francesco Zamponi for stimulating discussions. We also thank Sho Yaida for many of the ideas in Appendix \ref{sec:MMAppC} (notably how to sidestep MM's Appendix C, which very much depended on having $d=3$) and for identifying the formulae of Appendix \ref{sec:integrals_of_bessel_functions}. We acknowledge support from Canada's NSERC (Benoit Charbonneau), the National Science Foundation Grant no. NSF DMR-1055586 and from the Simons Foundation (\#454937, Patrick Charbonneau), and the National Science Foundation Grant no. DMR-1608086 (Grzegorz Szamel).
\end{acknowledgments}

\appendix
\section{Simplification of the Projected Force Correlation}
\label{sec:MMAppB}

Substituting the elements of Eq.~\eqref{eqn:restricted-ccomega} into into the right-hand side of Eq.~\eqref{eqn:XXavg} gives rise to an expression (Eq.~\eqref{eqn:XXfinal}) that involves the elements of the memory matrix $\RR$. In order to illustrate the analysis of this expression, we consider here the terms that involve its transverse elements, $\RR_{tt}$, \textit{i.e.}, 
\begin{widetext}
\begin{align}\label{eqn:A1}
\scp{\PF(\omega)\cdot \PF}_{tt} & = 
\frac{1}{(mV)^2}\sum_{\mathbf{K}_1,\mathbf{K}_2} \sum_{\vec{K}',\vec{K}''}   
\left[ \overline{W_l(\vec{K}_1)}\scp{l(\vec{K}_1)\cdot t(\vec{K}')}\RR_{tt}(\vec{K}',\vec{K}'',\omega)\scp{t(\vec{K}'')\cdot l(\vec{K}_2)}{W_l(\vec{K}_2)} \right.
\\ & \phantom{=\frac{1}{(mV)^2}\sum_{\mathbf{K}_1,\mathbf{K}_2} \sum_{\vec{K}',\vec{K}''}}
+ \left. \overline{W_t(\vec{K}_1)}\scp{t(\vec{K}_1)\cdot t(\vec{K}')}\RR_{tt}(\vec{K}',\vec{K}'',\omega)\scp{t(\vec{K}'')\cdot l(\vec{K}_2)}{W_l(\vec{K}_2)}
\right.
\nonumber\\ & \phantom{=\frac{1}{(mV)^2}\sum_{\mathbf{K}_1,\mathbf{K}_2} \sum_{\vec{K}',\vec{K}''}}
+ \left. \overline{W_l(\vec{K}_1)}\scp{l(\vec{K}_1)\cdot t(\vec{K}')}\RR_{tt}(\vec{K}',\vec{K}'',\omega)\scp{t(\vec{K}'')\cdot t(\vec{K}_2)}{W_t(\vec{K}_2)}
\right.
\nonumber\\ & \phantom{=\frac{1}{(mV)^2}\sum_{\mathbf{K}_1,\mathbf{K}_2} \sum_{\vec{K}',\vec{K}''}}
+ \left. \overline{W_l(\vec{K}_1)}\scp{t(\vec{K}_1)\cdot t(\vec{K}')}\RR_{tt}(\vec{K}',\vec{K}'',\omega)\scp{t(\vec{K}'')\cdot t(\vec{K}_2)}{W_t(\vec{K}_2)}
\right].\nonumber
\end{align} 

We first combine the longitudinal and transverse contributions in summations over $\vec{K}_1$ and $\vec{K}_2$, and then replace these summations by integrations over real space,
\begin{align}\label{eqn:A2}
\scp{\PF(\omega)\cdot \PF}_{tt} & = 
\frac{1}{m^2} \int \dd\vec{r}_1 \dd\vec{r}_2 \sum_{\vec{K}',\vec{K}''}   
\left[ \scp{[\overline{\vec{W}(\vec{r}_1)}\cdot\tilde{\vec{p}}(\vec{r}_1)]\cdot t(\vec{K}')}
\RR_{tt}(\vec{K}',\vec{K}'',\omega)\scp{t(\vec{K}'')\cdot [\tilde{\vec{p}}(\vec{r}_2)\cdot\vec{W}(\vec{r}_2)]}
\right],
\end{align}
\end{widetext}
where 
\begin{align}\label{eqn:defW}
\mathbf{W}(\mathbf{r}) =  \frac{\delta(r-\AAA)}{m\rho S_{d-1}r^{d-1} G(r)}
(\phi\mathbf{I}+\psi\hat{\mathbf{r}}\otimes\hat{\mathbf{r}}).
\end{align}
The momenta of different fluid particles being
uncorrelated, we can write
\begin{eqnarray}\label{eqn:pr1}
&& \left<\tilde{p}_{\alpha}(\mathbf{r}_1) \tilde{p}_{\gamma}(\mathbf{r}_2)\right>
\nonumber \\ && 
= \frac{m}{\beta} \delta_{\alpha\gamma} 
\left< \sum_i \delta[\mathbf{r}_1- (\mathbf{r}_i-\mathbf{R})] 
\delta[\mathbf{r}_2- (\mathbf{r}_i-\mathbf{R})]\right> 
\nonumber \\ && 
= \frac{m}{\beta} \delta_{\alpha\gamma} \delta(\mathbf{r}_1-\mathbf{r}_2)
\left< \sum_i \delta[\mathbf{r}_2- (\mathbf{r}_i-\mathbf{R})]\right>,
\end{eqnarray}
which implies that 
\begin{align}
 \frac{1}{m}\int \dd\mathbf{r}_1 
\langle[\overline{\vec{W}(\vec{r}_1)}\cdot&\tilde{\vec{p}}(\vec{r}_1)]\cdot t(\vec{K}')\rangle\notag
\\ & = (d-1) \frac{\rho G(\AAA)}{\beta} \overline{W_t(K')}.\label{eqn:tr}
\end{align}

\killinpaper{We use here that
\begin{align}
& \frac{1}{m}\int \dd\mathbf{r}_1 
\langle[\overline{\vec{W}(\vec{r}_1)}\cdot\tilde{\vec{p}}(\vec{r}_1)]\cdot t(\vec{K}')\rangle\notag
\\ & =
\frac{1}{m}\int \dd\mathbf{r}_1  
\langle[\overline{\vec{W}(\vec{r}_1)}\cdot\tilde{\vec{p}}(\vec{r}_1)]\cdot 
[\tilde{\vec{p}}(\vec{K}')\cdot (\vec{I}-\hat{\vec{K}'}\otimes \hat{\vec K}')]
\rangle\notag 
\\ & = \frac{1}{m}\int \dd\mathbf{r}_1  \int \dd\mathbf{r}_2 e^{-i\vec{K}'\cdot\vec{r}_2}
\overline{\vec{W}(\vec{r}_1)}: (\vec{I}-\hat{\vec{K}'}\otimes \hat{\vec K}') 
\notag 
\\ & \times \frac{m}{\beta} \delta(\mathbf{r}_1-\mathbf{r}_2)
\left< \sum_i \delta[\mathbf{r}_2- (\mathbf{r}_i-\mathbf{R})]\right>
\notag  
\\ & = \frac{1}{m} \int \dd\mathbf{r}_2 e^{-i\vec{K}'\cdot\vec{r}_2}
\overline{\vec{W}(\vec{r}_2)}: (\vec{I}-\hat{\vec{K}'}\otimes \hat{\vec K}') 
\notag 
\\ & \times \frac{m}{\beta} 
\left< \sum_i \delta[\mathbf{r}_2- (\mathbf{r}_i-\mathbf{R})]\right>\notag 
\end{align}

Since $W(r)\propto\delta(r-\AAA)$ and $\scp{\tilde n(\AAA)}=\rho G(\AAA)$, this last quantity is equal to
\begin{align}
\notag 
\\ & \int \dd\mathbf{r}_2 e^{-i\vec{K}'\cdot\vec{r}_2}
\overline{\vec{W}(\vec{r}_2)}: (\vec{I}-\hat{\vec{K}'}\otimes \hat{\vec K}') \frac{\rho G(\AAA)}{\beta} 
\\ & = \vec{W}(\vec{K}'): (\vec{I}-\hat{\vec{K}'}\otimes \hat{\vec K}') \frac{\rho G(\AAA)}{\beta}
\notag 
\\ & = (d-1) \frac{\rho G(\AAA)}{\beta} \overline{W_t(K')}
\end{align}
}

Finally, taking the thermodynamic limit to replace summation over wavevectors 
$\vec{K}'$ and $\vec{K}''$ by integrals, 
$\sum_{\mathbf{K}'} \to (V/(2\pi)^{d})\int \dd\vec{K}'$, gives the part of Eq.~\eqref{eqn:XXfinal}
that involves the transverse elements of the memory function, $\RR_{tt}$. Other parts can be obtained in a similar way.

\section{Generalized $d$ Viscosity Tensor}
\label{sec:etatensor}
The overarching message of Section 8 of Ref.~\onlinecite{Hansen:1986} is that in the hydrodynamic regime, for small wave-vectors and frequencies, correlations functions of macroscopic quantities are identical to the correlation functions of the corresponding microscopic quantities (an idea due to Onsager).  Sections 8.4 and 8.5 of Ref.~\onlinecite{Hansen:1986} respectively deal with the transverse current correlations and the longitudinal modes. In both cases, combinations of $\etashear $ and $\etabulk $ occur as limits $\omega\to0, \vec{K}\to0$ of macroscopic quantities, and thus of microscopic ones.

The key here is thus correctly identify the macroscopic tensor for arbitrary $d$.	
From a macroscopic perspective, the bulk and shear viscosity, $\etabulk $ and $\etashear $, appear in the stress tensor.  Classically in $d=3$, \killinpaper{following \S 15 of Ref.~\onlinecite{Landau:1987} or Eq.~(8.3.15) of Ref.~\onlinecite{Hansen:1986},} the macroscopic stress tensor is
\begin{equation*}
	\sigma_{ik}^\mathrm{macro} = -P\delta_{ik}+\etabulk  \delta_{ik}\frac{\p v_l}{\p x_l}+\etashear (\frac{\p v_i}{\p x_k}-\frac{\p v_k}{\p x_i}-\frac23\frac{\p v_l}{\p x_l}).
\end{equation*}
The factor $\frac23$ is there to make the tensor multiplying $\etashear $ trace-free.   In dimension $d$, this term is $\frac2d$.
In a coordinate-free language,  the macroscopic stress tensor is thus
\begin{widetext}
\begin{equation}\label{eqn:macroscopic stress tensor coordinate free}
	\vec{\sigma}^\mathrm{macro}(\vec{X},\vec{Y})=\Bigl(-P+\etabulk \div(\vec{v})\Bigr)\vec{X}\cdot \vec{Y}+\etashear \Bigl(\vec{X}\cdot \nabla_{\vec{Y}}\vec{v}+\vec{Y}\cdot \nabla_{\vec{X}}\vec{v}-\frac2d\div(\vec{v})\Bigr).
\end{equation}
\end{widetext}
(Note how this differs from Eq.~(33) of Ref.~\onlinecite{Charbonneau:2013} where $\etabulk $ was assumed to be zero.)
Following the arguments of Ref.~\onlinecite{Hansen:1986} yields
$\eta_{ijij}=\etashear $ for $i\neq j$ from the transverse current analysis and $\eta_{iiii}=\etabulk +\frac{2(d-1)}d\etashear $ from the longitudinal modes analysis. 

It thus remains to capture $\eta_{iijj}$ for $i\neq j$. To do so, we use the symmetries of the tensor $\vec{\eta}$. In an isotropic fluid, this tensor is invariant under rotation and reflection. As explained in Appendix A of Ref.~\onlinecite{Charbonneau:2013}, the first fundamental theorem of the orthogonal group impose the equation
\begin{equation}
	\eta_{ijkl} = c_1\delta_{ij}\delta_{kl}+c_2\delta_{ik}\delta_{jl}+c_3\delta_{il}\delta_{jk}.
\end{equation}
Thus, the only non-zero components are the $\eta_{iiii}$, $\eta_{iijj}$, and $\eta_{ijij}=\eta_{jiij}$. The fact that $\eta$ is symmetric forces $c_2=c_3=\etashear $. Given that $c_1+2\etashear =\eta_{iiii}=\etabulk +\frac{2(d-1)}d\etashear $, we have for $i\neq j$ that $\eta_{iijj}=c_1=\etabulk -\frac2d\etashear $. 

Note that Appendix A of Ref.~\onlinecite{Charbonneau:2013} commented that the linear viscosity $\eta_{\mathrm{L}}\equiv\eta_{iiii}$ is directly proportional to $\etashear$, and erroneously concluded that the numerical results of Ref.~\onlinecite{SHF95} suggesting otherwise likely reflected insufficient averaging. The mistake in this reasoning is that the pressure tensor is not \emph{instantaneously} tracefree, but only tracefree \emph{on average}. The bulk viscosity, $\etabulk$, thus does contribute to the linear viscosity, which validates numerical results of Ref.~\onlinecite{SHF95}.

\section{Drag on a fluid drop}
\label{sec:drag_on_a_drop_of_fluid}
We follow the ideas of Refs.~\onlinecite{Hadamard:1911,Rybczynski:1911} (recounted in Ref.~\onlinecite[\S 20, Problem 2]{Landau:1987}) to determine, in general dimension $d$, the drag coefficient, $\xi$, on a drop of fluid of viscosity $\eta_\mathrm{S}'$, pulled gravity with constant $g$, in a fluid with viscosity $\eta_\mathrm{S}$.   Note that we here adopt the notation and framework proposed in Section IV.A and Appendix F of Ref.~\onlinecite{Charbonneau:2013}. Note also taht a companion Maple 2017 code to compute the hydrodynamic drag on a drop of fluid suspended in a fluid with a different viscosity, in all dimensions, has been archived and can be accessed at \url{http://doi.org/10.7924/r4x061q6f}.

Consider a spherical fluid drop at rest, centered at the origin, within an outside fluid moving with a velocity field $\vec{v}$ that reaches a constant velocity $\vec{u}$ infinitely far away from the drop. As in Ref.~\onlinecite{Charbonneau:2013}, the velocity of the fluid outside the drop can be obtained from a potential function $f_\Out$ via
\begin{equation}
	\vec{v}_\Out=\vec{u}+\curl\ \curl (f_\Out\vec{u}).
\end{equation}
Similarly, a potential $f_\In$ inside the sphere yields
\begin{equation}
	\vec{v}_\In=\curl\ \curl (f_\In\vec{u}).
\end{equation}
Both $f_\In$ and $f_\Out$ are such that $\grad \Delta^2 f=0$, whence, for $d\neq 4$, we have
\begin{equation*}
	f=(-1)^d(\mathfrak{a}r^{4-d}+\mathfrak{b}r^{2-d}+\mathfrak{c}+\mathfrak{d}r^2+\mathfrak{e}r^4),
\end{equation*} 
while for $d=4$ we have $f=\mathfrak{a}\ln r+\mathfrak{b}r^{-2}+\mathfrak{c}+\mathfrak{d}r^2+\mathfrak{e}r^4$.  For the fluid outside the drop, we must have $\mathfrak{d}=\mathfrak{e}=0$ for the boundary velocity to be $\vec{u}$. For the fluid inside the sphere, we must have $\mathfrak{a}=\mathfrak{b}=0$ in order to prevent an unphysical singularity at the origin. In both cases, $\mathfrak{c}$ is immaterial because it disappears from the expression for $\vec{v}$. This potential determines both the pressure 
\begin{align}
	P&=P_0+(-1)^d\eta_\mathrm{S}\vec{u}\cdot\grad\Delta f,
\end{align}
and the components of the stress tensor 
\begin{multline}\label{eqn:sigma}
		\bm\sigma(\vec{X},\vec{Y})\equiv-P\vec{X}\cdot \vec{Y}+\eta_\mathrm{S}\Bigl(\vec{X}(\vec{Y}\cdot\vec{v})+\vec{Y}(\vec{X}\cdot\vec{v})\\
		-(\nabla_{\vec{X}}\vec{Y})\cdot \vec{v}-(\nabla_{\vec{Y}}\vec{X})\cdot \vec{v}\Bigr)
\end{multline}
in arbitrary directions $\vec{X},\vec{Y}$.

At the hydrodynamic radius, $\AAA$, which defines the interface between the two fluids, we have three boundary conditions:
\begin{itemize}
	\item the normal components $\vec{v}_\In|_{r=\AAA}\cdot\vec{n}$ and $\vec{v}_\Out|_{r=\AAA}\cdot\vec{n}$ must be $0$;
	\item $\vec{v}_\In|_{r=\AAA}=\vec{v}_\Out|_{r=\AAA}$;
	\item for directions $\vec{w}$ tangential to the sphere, $\bm\sigma_\In|_{r=\AAA}(\vec{n},\vec{w})=\bm\sigma_\Out|_{r=\AAA}(\vec{n},\vec{w})$.
\end{itemize}  
These conditions thus fix the constants:
\begin{align*}
	\mathfrak{a}&=\begin{cases}\frac{\AAA^{d-2}(d\eta_\mathrm{S}'+2\eta_\mathrm{S})}{2(d-1)(d-4)(\eta_\mathrm{S}+\eta_\mathrm{S}')},&\text{ if }d\neq 4\\
	      -\frac{(\eta_\mathrm{S}+2\eta_\mathrm{S}')\AAA^2}{3(\eta_\mathrm{S}+\eta_\mathrm{S}')},&\text{ if }d=4,\end{cases}\\
	\mathfrak{b}&=-\frac{\AAA^d\eta_\mathrm{S}'}{2(d-1)(\eta_\mathrm{S}+\eta_\mathrm{S}')},\\
	\mathfrak{d}&=-2\mathfrak{e}\AAA^2=\frac{(d-2)\eta_\mathrm{S}}{4(d-1)(\eta_\mathrm{S}+\eta_\mathrm{S}')}.
\end{align*}
Of these coefficients, only $\mathfrak{a}$ appears in the expression for the drag force, $\vec{F}=\xi\vec{u}$. Simplifying the above results, we obtain
\begin{equation}
	\xi_\mathrm{fluid} = \frac{4\eta_\mathrm{S}(2\eta_\mathrm{S}+d\eta_\mathrm{S}')\pi^{\frac d2}\AAA^{d-2}}{(d-1)\Gamma\bigl(\frac d2-1\bigr)(\eta_\mathrm{S}+\eta_\mathrm{S}')}.
\end{equation}

Equating this force to the gravitational pull gives the terminal speed, $V_\mathrm{t}$, of a slowly moving spherical drop of liquid of mass density $\varrho'$ (rather than number density) and viscosity $\eta_\mathrm{S}'$ through an ambient fluid of mass density $\varrho$ and viscosity $\eta_\mathrm{S}$,
\begin{equation}
 V_\mathrm{t} = \frac{(d-1)\AAA^2(\eta_\mathrm{S}+\eta_\mathrm{S}')g|\varrho-\varrho'|}{(d-2)d\eta_\mathrm{S}(2\eta_\mathrm{S}+d\eta_\mathrm{S}')}.
\end{equation}
This result thus generalizes the Hadamard--Rybczynski expression to all $d\geq3$\cite{Hadamard:1911,Rybczynski:1911}. 

If the drop of fluid has the same viscosity as the surrounding solvent, i.e., $\eta_\mathrm{S}=\eta_\mathrm{S}'$, we finally obtain
\begin{equation}
\xi_{\mathrm{fluid}}=\frac{(d^2-4)\eta_\mathrm{S} S_{d-1}\AAA^{d-2}}{2(d-1)}.
\end{equation}
Comparing this result with Eqs.~(44) and (48) from Ref.~\onlinecite{Charbonneau:2013}, we see that $\xi_\mathrm{fluid}$ is the mean of $\xi_\slip$ and $\xi_\stick$ in all $d\geq3$.

\section{Projected force self-correlation of the physical model}
\label{sec:MMAppC}
In this appendix, we compute the projected force self-correlation $\scp{\PF(\omega)\cdot \PF}$ under the structural hypotheses described in Sec.~\ref{sec:physical}.
 Note that fully computing $\RR\equiv\RRi^{-1}$ is here unnecessary; it suffices to compute a partial inverse defined as 
\begin{equation}\label{eqn:defb}
	\bb(\vec{K}\AAA)\equiv K^{d-1}\int\dd\vec{K}'\ \RR(\vec{K},\vec{K}')W_t(\vec{K}').
\end{equation}
It indeed follows that
\begin{align}
  \scp{\PF(\omega)\cdot \PF}\hide{= \cst\int_0^\infty\!\!\! \dd K\int\!\!\!\dd \hvec{K}\,	W_t(\vec{K})b(\vec{K}\AAA)\notag}
 &= \cst S_{d-1} \int_0^\infty\!\!\! \dd K\, W(K)\bb(K\AAA), 
  \label{eqn:IWb}	
\end{align}
where $\cst \equiv\bigl(\frac{(d-1)\rho G(\AAA)}{\beta}\bigr)^2(\frac{V}{(2\pi)^d})^2$.
In fact, even a full description of $\bb$ is extraneous. Because only $J_{\frac d2\pm 1}$ appear in $W_t$ (see Eq.~\eqref{eqn:WtK}), we only really need to compute the quantities
\begin{align}
	B_\pm\hide{\equiv \frac1{2A} \int_{-\infty}^\infty\!\! \dd K\ \bb(K)\frac{J_{\frac d2\pm1}(K)}{K^{\frac d2-1}}\notag}
	&\equiv\int_0^\infty\!\! \dd K \bb(K\AAA) \frac{J_{\frac d2\pm1}(K\AAA)}{(K\AAA)^{\frac d2-1}}\label{eqn:defBj},
\end{align}
for then
\begin{equation}\label{eqn:X2omega.X2}
	\scp{\PF(\omega)\cdot \PF}= \cst S_{d-1}\prefix \Bigl(\phi B_-+\frac{\psi}d(B_++B_-)\Bigr).
\end{equation}

Given that $\RRi(\vec{K}_1,\vec{K}_2)$ depends only on $\vec{K}_1\cdot \vec{K}_2$, $K_1$ and $K_2$, and that $W(\vec{K})$ only depends on $K$, it is natural to get rid of $\RRi$ and $\RR$ the dependence on direction .  
Let
\begin{align}
\widetilde{\RR}(K,K')&\equiv \frac{V}{(2\pi)^d}\iint\!\!\dd\hvec{K}\dd\hvec{K}'\, \RR(\vec{K},\vec{K}'),\label{eqn:tildeRR}\\
\widetilde{\RRi}(K,K')&\equiv \frac{V}{(2\pi)^d}\iint\!\!\dd\hvec{K}\dd\hvec{K}'\, \RRi(\vec{K},\vec{K}').\label{eqn:tildeRRi}
\end{align}
Given that 
\begin{equation*}
	\int_0^\infty\!\!\!\! \dd K' {K'}^{d-1}\widetilde{\RRi}(K,K')\widetilde{\RR}(K',K'')=S_{d-1}^2\frac{\delta(K-K'')}{K^{d-1}}
\end{equation*}  (see Eq.~\eqref{eqn:inversetilde}), we get  that 
\begin{align}
	W(K)\hide{=\int\!\!\!\dd K'' \delta(K-K'') W(K'')\notag}
	      \hide{=\frac{1}{S_{d-1}^2}\int\!\!\!\dd K'' K''^{d-1}\int \!\!\! \dd K' K'^{d-1}\widetilde{\RRi}(K,K')\widetilde{\RR}(K',K'')W(K'')\notag}
		  &=\frac1{S_{d-1}}\frac{V}{(2\pi)^d}\int_0^\infty\!\!\!\! \dd K' \widetilde{\RRi}(K,K')\bb(K'\AAA).\label{eqn:WintSb}
\end{align}
We have an explicit expression for $\RRi=\RRi_{tt}$: 
\begin{widetext}
\begin{align}
	\RRi(\vec{K},\vec{K}')&= -i\omega \scp{t(\vec{K})\cdot t(\vec{K}')}-\scp{\dot{t}(\vec{K})\cdot t(\vec{K}')}+[\dot{t}(\vec{K},\omega),\dot t(\vec{K}')]\notag\\
	&\approx -i\omega \scp{t(\vec{K})\cdot t(\vec{K}')}+\scp{\dot{t}(\vec{K},\omega=0)\cdot \dot{t}(\vec{K}')}\notag\\
	\hide{= \Bigl(-i\omega\frac m\beta\bigl(d-2+(\hvec{K}\cdot\hvec{K}')^2\bigr)+\frac{(\vec{K}\cdot \vec{K}')  \Bigl(d-3+2(\hvec{K}\cdot \hvec{K}')^2\Bigr)\eta_{\mathrm{S}}}{\rho\beta}\Bigr)[S(\vec{K}-\vec{K}')-1+N_\mathrm{f} \delta_{\vec{K},\vec{K}'}] \notag}
	&= \Bigl(-i\omega\frac m\beta\bigl(d-2+(\hvec{K}\cdot\hvec{K}')^2\bigr)+\frac{(\vec{K}\cdot \vec{K}')  \Bigl(d-3+2(\hvec{K}\cdot \hvec{K}')^2\Bigr)\eta_{\mathrm{S}}}{\rho\beta}\Bigr)\rho\int\dd\vec{r}G(r)e^{i(\vec{K}-\vec{K}')\cdot\vec{r}}.\label{eqn:explicit S}
\end{align}

To pass from $\RRi$ to $\widetilde\RRi$, it is convenient to introduce the quantities
\begin{align}
	\ttt_n(K,K',r)&\equiv \frac{1}{S_{d-1}}\iiint\!\!\dd\hvec{r}  \dd\hvec{K}\dd\hvec{K}'  (\hvec{K}\cdot \hvec{K}')^n e^{i(\vec{K}-\vec{K}')\cdot \vec{r}}\label{eqn:deftttn},\\
	\ttt_{0,2}(K,K',r)&\equiv (d-2)\ttt_0(K,K',r)+\ttt_2(K,K',r)\label{eqn:defttt02},\\
	\ttt_{1,3}(K,K',r)&\equiv (d-3)\ttt_1(K,K',r)+2\ttt_3(K,K',r)\label{eqn:defttt13},
\end{align}
which are all computed in Appendix \ref{sec:trig_integrals}. 
Then Eqs.~\eqref{eqn:tildeRRi} and \eqref{eqn:explicit S} yield
\begin{align}
	\widetilde{\RRi}(K,K')
	\hide{=\frac{V}{(2\pi)^d}\int\dd\hvec{K}\dd\hvec{K}' \RRi(\vec{K},\vec{K}')}
		\hide{=\frac{V}{(2\pi)^d}\int\dd\hvec{K}\dd\hvec{K}'\Bigl(-i\omega\frac m\beta\bigl(d-2+(\hvec{K}\cdot\hvec{K}')^2\bigr)+\frac{(\vec{K}\cdot \vec{K}')  \Bigl(d-3+2(\hvec{K}\cdot \hvec{K}')^2\Bigr)\eta_{\mathrm{S}}}{\rho\beta}\Bigr)\rho\int\dd\vec{r}G(r)e^{i(\vec{K}-\vec{K}')\cdot\vec{r}}}
		\hide{=S_{d-1}\frac{V}{(2\pi)^d}\int_0^\infty\!\! \dd r\, r^{d-1}\rho G(r) \Bigl(-i\omega \frac{m}{\beta}\bigl(\ttt_{0,2}(K,K',r)+\frac{KK'\eta_{\mathrm{S}}}{\rho\beta}\ttt_{1,3}(K,K',r)
		\Bigr)}
	&=\frac{S_{d-1}\etashear V}{\beta(2\pi)^d}\int_0^\infty\!\! \dd r\, r^{d-1} G(r) \Bigl(KK'\ttt_{1,3}(K,K',r)-\frac{\delta^2}{\AAA^2}\ttt_{0,2}(K,K',r)
	\Bigr).\label{eqn:RRi_simple}
\end{align}

While the case of interest is $G(r)=\Theta(r-\AAA)$, we momentarily consider 
\begin{equation}\label{eqn:defG}
	G(r)=\begin{cases} 1,&\text{ if }r\geq \AAA,\\
	     1-\epsilon, &\text{ if }r<\AAA,\end{cases}
\end{equation}
and let 
\begin{equation}
	\widetilde \RRi_q(K,K')\equiv  \frac{S_{d-1}\eta_{\mathrm{S}}V}{\beta (2\pi)^d}\int_0^q \!\dd r\, r^{d-1} \Bigl(
	KK'\ttt_{1,3}(K,K',r)-\frac{\delta^2}{\AAA^2}\ttt_{0,2}(K,K',r)
	\Bigr).
\end{equation}
We then have
\begin{equation}\label{eqn: RRi = RRi_infty - eps RRi_A}
	\widetilde \RRi(K,K') = \widetilde\RRi_\infty(K,K')-\epsilon \widetilde\RRi_\AAA(K,K').
\end{equation}
To integrate the first term in the above expression up to $q=\infty$, we use Hankel's theorem (see, e.g., Ref.~\onlinecite[Sec.~14.4]{Watson}), which states that for any $\nu>-\frac12$,
\begin{equation}\label{eqn:HankelThm}
	\int_0^\infty \!\!\dd r\,  r J_\nu(Kr)J_\nu(K'r) = \frac{\delta(K-K')}{K}.
\end{equation}
Thus \killinpaper{given that $\ttt_{i,j}=\frac{1}{r^{d-2}(KK')^{\frac d2-1}}\sum_{\nu} a_\nu^{i,j} J_\nu(Kr)J_\nu(K'r)$, we have that
\begin{align}
	\int_0^\infty \!\!\dd r\, r^{d-1} \ttt_{i,j}&=\frac{\delta(K-K')}{K^{d-1}}\sum_\nu a^{i,j}_\nu.
\end{align}
Given Eqns.~\eqref{eqn:RRi_simple},\eqref{eqn:t02}, and \eqref{eqn:t13}, we get}
\begin{align}
	\widetilde\RRi_\infty(K,K') &= \factorout d^2\bigl(K^2-\frac{\delta^2}{\AAA^2}\bigr)\frac{\delta(K-K')}{K^{d-1}},
\end{align}
with
$\factorout\equiv \frac{(d-1)S_{d-1}V\eta_{\mathrm{S}}}{d^2\beta }$.

Computing the second term is somewhat more involved. For notational simplicity, let 
\begin{equation}
	J_\pm(K) \equiv \frac{J_{\frac d2\pm1}(K\AAA)}{(K\AAA)^{\frac d2-1}},
\end{equation}
and let $\coeff_-$ and $\coeff_+$ be such that
\begin{equation}
	\frac{S_{d-1}(2\pi)^d}{V\factorout} W(K) = \coeff_-J_-(K)+\coeff_+J_+(K).
\end{equation}
A clever use of identities of Bessel functions and their integrals (outlined in Appendix \ref{sec:integration_over_0_a}) can then be used to obtain a simplified expression that only uses $J_{\frac d2\pm1}$:
\begin{align}
	\widetilde\RRi_\AAA(K,K')&=\factorout\AAA^{d-2}
	\Biggl(
	(\delta^2-2d^2)
J_{+}(K)J_{+}(K')
	-(d-1)\delta^2 
J_{-}(K)J_{-}(K')\notag\\
	&
	\phantom{=\factorout\Biggl(}
	+\bigl(\delta^2+d\frac{K^2((\AAA K')^2-\delta^2)}{K^2-(K')^2}\bigr) J_{+}(K)J_{-}(K')
	+\bigl(\delta^2-d\frac{(K')^2((\AAA K)^2-\delta^2)}{K^2-(K')^2}\bigr) J_{-}(K)J_{+}(K')
	\Biggr).\label{eqn:SA in text}
\end{align}

From Eqs.~\eqref{eqn:WintSb} and \eqref{eqn: RRi = RRi_infty - eps RRi_A}, we have
\begin{align}
	\frac{S_{d-1}(2\pi)^d}{V} W(K)
		\hide{=\int_0^\infty \dd K'\  \widetilde{\RRi}(K,K')\bb(K'\AAA)\notag}
		\hide{=\int_0^\infty \dd K'\ \widetilde{\RRi}_\infty(K,K')\bb(K'\AAA)
		-\epsilon \int_0^\infty \dd K'\  \widetilde{\RRi}_\AAA(K,K')\bb(K'\AAA)\notag}
		&=\factorout d^2(K^2-\frac{\delta^2}{\AAA^2}) \frac{\bb(K\AAA)}{K^{d-1}}
-\epsilon \int_0^\infty \dd K'\  \widetilde{\RRi}_\AAA(K,K')\bb(K'\AAA)\label{eqn:W=b+int},
\end{align}
with
\begin{align}
	\int_0^\infty \!\!\!\! \dd K'\, \widetilde \RRi_\AAA(K,K')\bb(K'\AAA)   &= \factorout \AAA^{d-2}
		\Biggl(
		(\delta^2-2d^2)B_+J_{+}(K)
		-(d-1)\delta^2B_- J_{-}(K)
		+\delta^2B_+ J_{-}(K)\notag\\
		&\phantom{=\factorout \Biggl(}
		+\delta^2 B_-J_{+}(K)
		-K^2d\ J_{+}(K)\int_0^\infty\!\!\!\! \dd K'\,  \frac{\bigl((\AAA K')^2-\delta^2\bigr)}{K'^2-K^2} J_{-}(K')\bb(K'\AAA)\notag	\\
		&\phantom{=\factorout \Biggl(}
		+\bigl((\AAA K)^2-\delta^2\bigr)d\ J_{-}(K)\int_0^\infty\!\!\!\! \dd K'\,  \frac{K'^2}{K'^2-K^2} J_{+}(K')	\bb(K'\AAA)\Biggr).\label{eqn:int b(K1) S1}
\end{align}

Let 
\begin{align}
	\tilde\coeff_-&\equiv \coeff_-+\epsilon\AAA^{d-2}\bigl(\delta^2B_+-(d-1)\delta^2B_-\bigr), \\
	\tilde\coeff_+&\equiv \coeff_++\epsilon\AAA^{d-2}\bigl((\delta^2-2d^2)B_++\delta^2B_-\bigr).
\end{align}
Then using Eqns.~\eqref{eqn:W=b+int} and \eqref{eqn:int b(K1) S1}, we find
\begin{align}
	\tilde\coeff_-J_-(K)+\tilde\coeff_+J_+(K)&=\frac{d^2}{\AAA^2}((\AAA K)^2-\delta^2)\frac{\bb(K\AAA)}{K^{d-1}}\notag \\
	&\phantom{=}+\epsilon K^2d\ J_{+}(K)\AAA^{d-2}\int_0^\infty\!\!\!\! \dd K'\, \frac{\bigl((\AAA K')^2-\delta^2\bigr)}{K'^2-K^2} J_{-}(K')\bb(K'\AAA)\notag	\\
	&\phantom{=}-\epsilon\bigl((\AAA K)^2-\delta^2\bigr)d\ J_{-}(K)\AAA^{d-2}\int_0^\infty\!\!\!\! \dd K'\,  \frac{K'^2}{K'^2-K^2} J_{+}(K')	\bb(K'\AAA),
\end{align}
hence
\begin{align}
	\AAA^{d-3}\frac{\bb(K\AAA)}{(\AAA K)^{d-1}}
	\hide{=\frac1{d^2}\frac{\tilde\coeff_-J_-(K)+\tilde\coeff_+J_+(K)}{(\AAA K)^2-\delta^2}\notag}
	\hide{\phantom{=}-\frac{\epsilon}d \frac{K^2J_{+}(K)}{(\AAA K)^2-\delta^2}\AAA^{d-2}\int_0^\infty\!\!\!\! \dd K'\, \frac{\bigl((\AAA K')^2-\delta^2\bigr)}{K'^2-K^2} J_{-}(K')\bb(K'\AAA)\notag	}
	\hide{\phantom{=}+\frac{\epsilon}d J_{-}(K)\AAA^{d-2}\int_0^\infty\!\!\!\! \dd K'\,  \frac{K'^2}{K'^2-K^2} J_{+}(K')	\bb(K'\AAA)\notag}
	%
	%
	\hide{=\frac1{d^2}\frac{\tilde\coeff_-J_-(K)+\tilde\coeff_+J_+(K)}{(\AAA K)^2-\delta^2}\notag}
    \hide{\phantom{=}+\frac{\epsilon}d J_{+}(K)\AAA^{d-2}\int_0^\infty\!\!\!\! \dd K'\, \Bigl(\frac{K'^2}{K^2-K'^2}-\frac{\delta^2}{(\AAA K)^2-\delta^2}\Bigr) J_{-}(K')\bb(K'\AAA)\notag}
	\hide{\phantom{=}+\frac{\epsilon}d J_{-}(K)\AAA^{d-2}\int_0^\infty\!\!\!\! \dd K'\,  \frac{K'^2}{K'^2-K^2} J_{+}(K')	\bb(K'\AAA)	\notag}
	%
	%
	\hide{=\frac1{d^2}\frac{\tilde\coeff_-J_-(K)+\tilde\coeff_+J_+(K)}{(\AAA K)^2-\delta^2}
		-\frac{\epsilon}d J_{+}(K)\frac{\delta^2}{(\AAA K)^2-\delta^2} \AAA^{d-2}\int_0^\infty\!\!\!\! \dd K'\, J_{-}(K')\bb(K'\AAA)\notag}
	\hide{\phantom{=}+\frac{\epsilon}d \AAA^{d-2}\int_0^\infty\!\!\!\! \dd K'\, \frac{K'^2}{K^2-K'^2} \bigl(J_{+}(K)J_{-}(K')-J_{-}(K)J_{+}(K')\bigr)\bb(K'\AAA)\notag}
			%
			%
				&=\frac1{d^2}\frac{\tilde\coeff_-J_-(K)+(\tilde\coeff_+-\AAA^{d-2}B_-\epsilon d \delta^2)J_+(K)}{(\AAA K)^2-\delta^2}\notag\\
					&\phantom{=}+\frac{\epsilon}d \AAA^{d-2}\int_0^\infty\!\!\!\! \dd K'\, \frac{K'^2}{K^2-K'^2} \Bigl(J_{+}(K)J_{-}(K')-J_{-}(K)J_{+}(K')\Bigr)\bb(K'\AAA)	
	.\label{eqn:Master Eqn}
\end{align}
Multiplying this last equation by $(\AAA K)^{\frac d2}J_{\frac d2\pm1}(K\AAA)$ and integrating (using the identities of Appendix \ref{sec:integrals_of_bessel_functions}) gives
\begin{align}\label{eqn:B+- identities}
	\AAA^{d-3}B_{\pm}&=\frac{i\pi}{2\AAA d^2}\Bigl(\tilde\coeff_-H^{(1)}_{\frac d2-1}(\delta)J_{\frac d2\pm 1}(\delta)+(\tilde\coeff_+-\AAA^{d-2}B_-\epsilon d \delta^2)H^{(1)}_{\frac d2\pm1}(\delta)J_{\frac d2+ 1}(\delta)\Bigr)+\delta_{+,\pm}\epsilon \AAA^{d-3}B_+.
\end{align}
\killinpaper{(The last line yields
\begin{align*}
	\frac{\epsilon}d\AAA^{d-2} \int_0^\infty& \!\!\!\! \dd K\int_0^\infty\!\!\!\! \dd K_2 (K\AAA)J_{\frac d2\pm1}(K\AAA) \frac{K_2^2}{K^2-K_2^2} \Bigl(J_{\frac d2+1}(K\AAA)J_{-}(K_2)-J_{\frac d2-1}(K\AAA)J_{+}(K_2)\Bigr)\bb(K_2\AAA)\\
	&=\frac\epsilon d \AAA^{d-2}\int_0^\infty\!\!\!\! \dd K_2 \frac{(\AAA K_2)^2}{(K_2\AAA)^{\frac d2-1}}\bb(K_2\AAA)\int_0^\infty \!\!\!\! \dd K \frac{(K\AAA)J_{\frac d2\pm1}(K\AAA)\Bigl(J_{\frac d2+1}(K\AAA)J_{\frac d2-1}(K_2\AAA)-J_{\frac d2-1}(K\AAA)J_{\frac d2+1}(K_2\AAA)\Bigr)}{(\AAA K)^2-(\AAA K_2)^2}\\
	&=\delta_{\pm+}\frac\epsilon\AAA\AAA^{d-2}\int_0^\infty \dd K_2 \bb(K_2\AAA)\frac{J_{\frac d2+1}(K_2\AAA)}{(K_2\AAA)^{\frac d2-1}}=\delta_{\pm+}\epsilon\AAA^{d-3} B_+
\end{align*})}
Setting $\epsilon=1$ and substituting in Eq.~\eqref{eqn:X2omega.X2} finally gives
\begin{equation}
	\scp{\PF(\omega)\cdot \PF} = \frac{(d-1)}{\AAA^{d-2}\beta\etashear S_{d-1}\delta^2}\frac{\Bigl(\bigl(d^3-\frac{d\delta^2}{2}\bigr)\phi^2+2d^2\phi\psi+\bigl(d+\frac{\delta^2}{2})\psi^2\Bigr)H^{(1)}_{\frac d2-1}(\delta)+\frac{\delta^2\psi^2}2H^{(1)}_{\frac d2+1}(\delta)}%
{\bigl(d^2-d-\frac{\delta^2}2\bigr)H^{(1)}_{\frac d2-1}(\delta)+\bigl(d^2-\frac{\delta^2}2\bigr)H^{(1)}_{\frac d2+1}(\delta)}.
\end{equation}
\end{widetext}


\section{The averaging trick} 
\label{sec:integrating_out_the_spherical_component}
We describe here a particular useful trick that helps us understand the spherical integral of the Dirac delta function and simplify the computation of $\RR=\RRi^{-1}$. Recall that the relation between $\RR$ or $\RRi$ is either
\begin{equation}
	\sum_{\vec{K}_2}\RRi(\vec{K}_1,\vec{K}_2)\RR(\vec{K}_2,\vec{K}_3)=\delta_{\vec{K}_1-\vec{K}_3}
\end{equation}
in discrete space or
\begin{equation}
	\bigl(\frac{V}{(2\pi)^d}\bigr)^2\int\!\!\!\dd \vec{K}_2 \RRi(\vec{K}_1,\vec{K}_2)\RR(\vec{K}_2,\vec{K}_3)=\delta(\vec{K}_1-\vec{K}_3)
\end{equation}
in continuous space. Note that $\RRi(\vec{K}_1,\vec{K}_2)$ depends on $K_1,K_2$, $\vec{K}_1\cdot \vec{K}_2$, and on $\|\vec{K}_1-\vec{K}_2\|=\sqrt{K_1^2-K_2^2-2\vec{K}_1\cdot \vec{K}_2}$. In words, $\RRi(\vec{K}_1,\vec{K}_2)$, and thus $\RR(\vec{K}_1,\vec{K}_2)$, only depends  on the length of the vectors and the angle between them.

The averaging trick is as follows.  Let $c_d$ denote the volume of the rotation group $\SO(d)$. If $f$ is a function on $\mathbb{S}^{d-1}$, then 
\begin{equation}\label{eqn:averaging}
	\int_{\SSS^{d-1}}\!\!\!\!\!\!\!\!\! \dd \Omega  f(\Omega)= \frac1{c_{d-1}}\int_{\SO(d)}\!\!\!\!\!\!\!\!\! \dd g f(g\vec{e}_1).
\end{equation}
In particular, $S_{d-1}=\frac{c_{d}}{c_{d-1}}$ and the constants $c_d$ can  be computed recursively, with $c_1\equiv 1$ and 
\begin{equation}
	c_d\equiv S_{d-1}c_{d-1}.\label{eqn:defc}
\end{equation}

From Eqs.~\eqref{eqn:tildeRR} and \eqref{eqn:tildeRRi}, we then get
\begin{align}
	\widetilde{\RR}(K_2,K_3)
	\hide{= \frac{V}{(2\pi)^d}\iint\!\!\dd\hvec{K}_2d\hvec{K}_3\ \RR(\vec{K}_2,\vec{K}_3)\notag}
	&=\frac1{c_{d-1}^2}\frac{V}{(2\pi)^d}\iint \dd g_2\dd g_3\ \RR(K_2g_2\vec{e}_1,K_3g_3\vec{e}_1)\notag\\
	\hide{=\frac1{c_{d-1}^2}\frac{V}{(2\pi)^d}\iint \dd g_2\dd g_3\ \RR(K_2\vec{e}_1,K_3g_2^{-1}g_3\vec{e}_1)\notag}
	&=\frac1{c_{d-1}^2}\frac{V}{(2\pi)^d}\iint \dd g_2d\tilde{g}_3\ \RR(K_2\vec{e}_1,K_3\tilde{g}_3\vec{e}_1)\notag\\
	\hide{=\frac{c_d}{c_{d-1}^2}\frac{V}{(2\pi)^d}\int d\tilde{g}_3\ \RR(K_2\vec{e}_1,K_3\tilde{g}_3\vec{e}_1)\notag}
	&=S_{d-1}\frac{V}{(2\pi)^d}\int\dd\hvec{K}_3\ \RR(K_2\vec{e}_1,\vec{K}_3),
\end{align}
and similarly
\begin{align}
	\widetilde{\RRi}(K_1,K_2)
	\hide{\frac{V}{(2\pi)^d}\iint\!\!\dd\hvec{K}_2d\hvec{K}_3\ \RRi(\vec{K}_2,\vec{K}_3)\notag}
	&=S_{d-1}\frac{V}{(2\pi)^d}\int\dd\hvec{K}_2\ \RRi(\vec{K}_2,K_3\vec{e}_1).\label{eqn:RRisingleintegral}
\end{align}
Hence\begin{widetext}
\begin{align}
	\iiint\!\!\! \dd\hvec{K}_1 \dd\vec{K}_2 \dd\hvec{K}_3 \RRi(\vec{K}_1,\vec{K}_2)\RR(\vec{K}_2,\vec{K}_3)
	\hide{=\frac1{c_{d-1}^3}\int\!\!\!\dd K_2 K_2^{d-1}\!\! \iiint\!\!\! \dd g_1\dd g_2\dd g_3 \RRi(K_1g_1\vec{e}_1,K_2g_2\vec{e}_1)\RR(K_2g_2\vec{e}_1,K_3g_3\vec{e}_1)\notag}
	\hide{=\frac1{c_{d-1}^3}\int \!\!\! \dd K_2 K_2^{d-1}\!\! \iiint \dd g_1\dd g_2\dd g_3 \RRi(K_1g_2^{-1}g_1\vec{e}_1,K_2\vec{e}_1)\RR(K_2\vec{e}_1,K_3g_2^{-1}g_3\vec{e}_1)\notag}
	\hide{=\frac1{c_{d-1}^3}\int \!\!\! \dd K_2 K_2^{d-1}\!\! \iiint d\tilde g_1\dd g_2d\tilde g_3 \RRi(K_1\tilde g_1\vec{e}_1,K_2\vec{e}_1)\RR(K_2\vec{e}_1,K_3\tilde g_3\vec{e}_1)\notag}
	&=\frac{c_d}{c_{d-1}^3}\int \!\!\! \dd K_2 K_2^{d-1}\!\! \iint d\tilde g_1d\tilde g_3 \RRi(K_1\tilde g_1\vec{e}_1,K_2\vec{e}_1)\RR(K_2\vec{e}_1,K_3\tilde g_3\vec{e}_1)\notag\\
	\hide{=\frac{c_d}{c_{d-1}}\int \!\!\! \dd K_2 K_2^{d-1} \iint \dd\hvec{K}_1d\hvec{K}_3 \RRi(\vec{K}_1,K_2\vec{e}_1)\RR(K_2\vec{e}_1,\vec{K}_3)\notag}
	&=\frac1{S_{d-1}}\bigl(\frac{V}{(2\pi)^d}\bigr)^{-2}\int \!\!\! \dd K_2\ K_2^{d-1} \widetilde{\RRi}(K_1,K_2)\widetilde{\RR}(K_2,K_3).\label{eqn:intK2RRi}
\end{align}

Now, suppose $f$ is invariant under rotation. Then
\begin{align}
	f(K_1)\hide{=f(\vec{K}_1) \notag}
	&= \int\!\!\!\dd \vec{K}_2 \delta(\vec{K}_1-\vec{K}_2)f(\vec{K}_2)\notag\\
 &=\int_0^\infty\!\!\!\dd K_2K_2^{d-1}f(K_2)\!\int\!\!\!\dd \hvec{K}_2\delta(\vec{K}_1-\vec{K}_2)\notag,
\end{align}
and hence 
\begin{equation}\label{eqn:intdelta}
	\iint\!\!\! \dd\hvec{K}\dd\hvec{K}'\ \delta(\vec{K}-\vec{K}') = S_{d-1}\frac{\delta(K-K')}{K^{d-1}}.
\end{equation}

Therefore, combining Eqs.~\eqref{eqn:intK2RRi} and \eqref{eqn:intdelta}, we get
\begin{align}
\int_0^\infty\!\!\! \dd K_2 \ K_2^{d-1}\widetilde{\RRi}(K_1,K_2)\widetilde{\RR}(K_2,K_3)
&=S_{d-1}\bigl(\frac{V}{(2\pi)^d}\bigr)^2\iiint \!\!\dd\hvec{K}_1\dd\vec{K}_2\dd\hvec{K}_3	\RRi(\vec{K}_1,\vec{K}_2)\RR(\vec{K}_2,\vec{K}_3)\notag\\
&=S_{d-1}\iint\!\! \dd\hvec{K}_1\dd\hvec{K}_3\ \delta(\vec{K}_1-\vec{K}_3)\notag\\
&=S_{d-1}^2\frac{\delta(K_1-K_3)}{K_3^{d-1}}.\label{eqn:inversetilde}
\end{align}
\end{widetext}

\section{Trigonometric integrals}
\label{sec:trig_integrals}
This Appendix collects the evaluation of various integrals whose value can be expressed in terms of Bessel functions. 

First, we consider simple trigonometric integrals
\begin{align}
s_n&\equiv  \int_0^\pi\!\!\! \dd\phi\, \sin^n\phi =\frac{S_{n+1}}{S_n},\\
s_{n,m}&\equiv \int_0^\pi\!\!\!	\dd\phi\, \sin^n\phi \cos^m\phi\notag\\
&=\begin{cases} 0,&\text{ if $m$ is odd},\\
   \sum_{k=0}^{\frac m2}(-1)^k\binom{\frac m2}k s_{n+2k},&\text{ if $m$ is even}.
\end{cases}
\end{align}
Using that $S_d=\frac{2\pi}{d-1}S_{d-2}$, we get the relation
\begin{equation}
	s_n=\frac{S_{n+1}}{S_n}=\frac{\frac{2\pi}nS_{n-1}}{\frac{2\pi}{n-1}S_{n-2}}= \frac{n-1}{n}s_{n-2},
\end{equation}
and thus
\begin{align}
	s_{d-3,2}&=\frac{1}{d-1}\frac{S_{d-2}}{S_{d-3}},\\
	s_{d-3,4}&=\frac{3}{d^2-1}\frac{S_{d-2}}{S_{d-3}}.
\end{align}

Second, let
\begin{align}\label{eqn:defvarsigma}
	\varsigma_n(x)&\equiv\int_0^\pi \dd\theta\ \sin^{n}(\theta) e^{ix\cos(\theta)},\\
	\varsigma_{n,m}(x)&\equiv\int_0^\pi \dd\theta\ \sin^{n}(\theta)\cos^{m}(\theta) e^{ix\cos(\theta)}.
\end{align}
As \killinpaper{$\varsigma''_{n}=-\varsigma_{n,2}=-\varsigma_n+\varsigma_{n+2}$,
we have
\begin{equation}
	\varsigma''_n+\varsigma_n = \varsigma_{n+2}= -i\frac{(n+1)}{x}\varsigma_{n,1}=-(n+1)\frac{\varsigma_n'}{x},
\end{equation}
or equivalently}	$x^2\varsigma_n'' +(n+1)x\varsigma_n'+x^2\varsigma_n=0$, we have\killinpaper{, as solutions to this differential equation lie in the span of $x^{-\frac n2}J_{\frac n2}(x)$ and $x^{-\frac n2}Y_{\frac n2}(x)$. Comparing $\varsigma_n(0)=s_n$ and $\lim_{x\to 0}x^{-\frac n2}J_{\frac n2}(x)$ (and noting that $\lim_{x\to 0}x^{-\frac n2}Y_{\frac n2}(x)$ is not finite), we find that}
\begin{align}
	\varsigma_n(x) \hide{= 2^{\frac n2}\sqrt{\pi}\Gamma(\frac{n+1}2) \frac{J_{\frac n2}(x)}{x^{\frac n2}}\notag}
	&=\frac{(2\pi)^{\frac n2+1}}{S_n}\frac{J_{\frac n2}(x)}{x^{\frac n2}}.
\end{align}
Using integrating by parts\killinpaper{ to get a better hold of $\varsigma_{n,1}$.  Indeed,  if one chooses $u=\sin^n(\theta)e^{ix\cos(\theta)}$ and $dv=\dd\theta \cos(\theta)$, one gets $v=\sin(\theta)$ and $du=\dd\theta (n\sin^{n-1}\theta\cos\theta e^{ix\cos(\theta)}-ix\sin^{n+1}\theta e^{ix\cos(\theta)})$.  When $n>1$, $u(0)=u(\pi)=0$, hence
$	\varsigma_{n,1}=-n\varsigma_{n,1}+ix\varsigma_{n+2}$,
that is } we find
\begin{equation}
	\varsigma_{n,1}=\frac{ix}{n+1}\varsigma_{n+2}.
\end{equation}
Thus 
\begin{align}
	\varsigma_{n,2m}&=\sum_{k=0}^m(-1)^k\binom{m}{k}\varsigma_{n+2k},\\
	\varsigma_{n,2m+1}
	\hide{=\sum_{k=0}^m(-1)^k\binom{m}{k}\varsigma_{n+2k,1}\notag}
	&=\sum_{k=0}^m(-1)^k\binom{m}{k}\frac{ix}{n+2k+1}\varsigma_{n+2k+2} .
\end{align}

We use the averaging trick of Eq.~\eqref{eqn:averaging} to compute the quantities $\ttt_n(K,K',r)$. For the lowest order, we get
\begin{widetext}
\begin{align}
	\ttt_0(R,R',r)&= \frac{1}{S_{d-1}} \int_{\SSS^{d-1}}\dd \hvec{r}  \ \int_{\SSS^{d-1}}\dd \hvec{K} \int_{\SSS^{d-1}}\dd \hvec{K}' \ e^{iKr(\hvec{K}\cdot \hvec{r})-iK'r(\hvec{K}'\cdot \hvec{r})}\notag\\
	\hide{=\frac{1}{S_{d-1}}\frac{1}{c_{d-1}}\int_{SO(d)}\dd g\int_{\SSS^{d-1}}\dd \hvec{K} \int_{\SSS^{d-1}}\dd \hvec{K}' \ e^{iKr(\hvec{K}\cdot g\vec{e}_1)-iK'r(\hvec{K}'\cdot g\vec{e}_1)}\notag}
	\hide{=\frac{1}{S_{d-1}}\frac{1}{c_{d-1}}\int_{SO(d)}\dd g\int_{\SSS^{d-1}}\dd \hvec{K} \int_{\SSS^{d-1}}\dd \hvec{K}' \ e^{iKr(g^{-1}\hvec{K}\cdot \vec{e}_1)-iK'r(g^{-1}\hvec{K}'\cdot \vec{e}_1)}\notag}
	&=\frac{1}{S_{d-1}}\frac{1}{c_{d-1}}\int_{SO(d)}\dd g\int_{\SSS^{d-1}}\dd \hvec{K} \int_{\SSS^{d-1}}\dd \hvec{K}' \ e^{iKr(\hvec{K}\cdot \vec{e}_1)-iK'r(\hvec{K}'\cdot \vec{e}_1)}\notag\\
		\hide{=\frac{1}{S_{d-1}}\int_{\SSS^{d-1}}\dd \hvec{r} \int_{\SSS^{d-1}}\dd \hvec{K} \int_{\SSS^{d-1}}\dd \hvec{K}' \ e^{iKr(\hvec{K}\cdot \vec{e}_1)-iK'r(\hvec{K}'\cdot \vec{e}_1)}\notag}
	\hide{=  \Bigl(\int_{\SSS^{d-1}}\dd \hvec{K}\ e^{iKr(\hvec{K}\cdot \vec{e}_1)} \Bigr)\Bigl(\int_{\SSS^{d-1}}\dd \hvec{K}' \ e^{-iK'r(\hvec{K}'\cdot \vec{e}_1)}\Bigr)\notag}
	&=  \Bigl(\int_{\SSS^{d-1}}\dd \hvec{K}\ e^{iKr(\hvec{K}\cdot \vec{e}_1)} \Bigr)\Bigl(\int_{\SSS^{d-1}}\dd \hvec{K}' \ e^{iK'r(\hvec{K}'\cdot \vec{e}_1)}\Bigr).
\end{align}
We can then use spherical coordinates (with $\hvec{K}=\cos(\theta)\vec{e}_1+\sin(\theta)\hvec{f}$ and $\hvec{f}\in \SSS^{d-2}$) to find that
  \begin{align*}
  	\int_{\SSS^{d-1}}\!\!\!\!\!\! \dd \hvec{K} e^{iKr(\hvec{K}\cdot \vec{e}_1)} &= \int_0^\pi\!\!\!  \dd\theta \!\!\int_{\SSS^{d-2}}\!\!\!\!\! \dd \hvec{f} \sin^{d-2}(\theta) e^{iKr\cos(\theta)}\\
	&=S_{d-2}\!\!\int_0^\pi\!\!\!  \dd\theta \sin^{d-2}(\theta) e^{iKr\cos(\theta)},
  \end{align*}
and thus that
\begin{align*}
	\ttt_0(K,K',r)&=S_{d-2}^2\varsigma_{d-2}(Kr)\varsigma_{d-2}(K'r).
\end{align*}
In general, we get
\begin{align}
	\ttt_n(K,K',r)&= \frac{1}{S_{d-1}}\int_{\SSS^{d-1}}\!\!\!\!\!\! \dd \hvec{r} \int_{\SSS^{d-1}}\!\!\!\!\!\! \dd \hvec{K}  \int_{\SSS^{d-1}}\!\!\!\!\!\! \dd \hvec{K}'  (\hvec{K}\cdot\hvec{K}')^ne^{iKr(\hvec{K}\cdot \hvec{r})-iK'r(\hvec{K}'\cdot \hvec{r})}\notag \\
	&=\int_{\SSS^{d-1}}\!\!\!\!\!\! \dd \hvec{K}  \int_{\SSS^{d-1}}\!\!\!\!\!\! \dd \hvec{K}'  (\hvec{K}\cdot\hvec{K}')^ne^{iKr(\hvec{K}\cdot \vec{e}_1)-iK'r(\hvec{K}'\cdot \vec{e}_1)}\notag\\
	&=(-1)^n\int_{\SSS^{d-1}}\!\!\!\!\!\! \dd \hvec{K}  \int_{\SSS^{d-1}}\!\!\!\!\!\! \dd \hvec{K}'  (\hvec{K}\cdot\hvec{K}')^ne^{iKr(\hvec{K}\cdot \vec{e}_1)+iK'r(\hvec{K}'\cdot \vec{e}_1)},
\end{align}
which, using spherical coordinates,
\begin{align}
	\hvec{K}&=\cos(\theta)\vec{e}_1+\sin(\theta)\hvec{f}, \text{ with }\hvec{f}\in \SSS^{d-2},\notag\\
	\hvec{K}'&=\cos(\theta')\vec{e}_1+\sin(\theta')\cos(\phi')\hvec{f}+\sin(\theta')\sin(\phi')\hvec{f}',\text{ with }\hvec{f}'\in \SSS^{d-3}\notag
\end{align}
gives
\begin{align}
	\ttt_n(K,K',r)&=(-1)^n\int_{\SSS^{d-2}}\!\!\!\!\!\!\! \dd \hvec{f}\int_0^\pi\!\!\! \dd\theta \sin^{d-2}\theta\!\int_{\SSS^{d-3}}\!\!\!\!\!\!\!\dd \hvec{f}'\!\int_0^\pi\!\!\! \dd\theta'\notag\\
	&\phantom{(-1)^n\int_{\SSS^{d-2}}\!\!\!\!\!\!\! \dd \hvec{f}}
	\int_0^\pi \!\!\!\dd\phi' \sin^{d-2}\theta'\sin^{d-3}\phi' e^{iKr\cos\theta+iK'r\cos\theta'}(\cos\theta\cos\theta'+\sin\theta\sin\theta'\cos\phi')^n\notag\\
	\hide{=(-1)^nS_{d-2}S_{d-3}\int_0^\pi\!\!\! \dd\theta \int_0^\pi\!\!\! \dd\theta'\int_0^\pi \!\!\!\dd\phi' \sin^{d-2}\theta\sin^{d-2}\theta'\sin^{d-3}\phi' e^{iKr\cos\theta+iK'r\cos\theta'}(\cos\theta\cos\theta'+\sin\theta\sin\theta'\cos\phi')^n\notag}
	\hide{=(-1)^n\sum_{k=0}^n \binom{n}{k}S_{d-2}S_{d-3}\int_0^\pi\!\!\! \dd\theta \int_0^\pi\!\!\! \dd\theta'\int_0^\pi \!\!\!\dd\phi' \sin^{d-2+n-k}\theta\cos^k\theta\sin^{d-2+n-k}\theta'\cos^k\theta'\sin^{d-3}\phi'\cos^{n-k}\phi' e^{iKr\cos\theta+iK'r\cos\theta'}\notag}
	&=(-1)^n\sum_{k=0}^n \binom{n}{k}S_{d-2}S_{d-3} s_{d-3,n-k}\varsigma_{d-2+n-k,k}(Kr)\varsigma_{d-2+n-k,k}(K'r).
\end{align}
Expanding these formulas gives
\begin{align*}
	\ttt_0(K,K',r)&=\frac{(2\pi)^{d}}{r^{d-2}(KK')^{\frac d2-1}}J_{\frac {d}2-1}(Kr)J_{\frac {d}2-1}(K'r),\\
	\ttt_1(K,K',r)&= \frac{(2\pi)^{d}}{r^{d-2}(KK')^{\frac d2-1}}J_{\frac d2}(Kr)J_{\frac d2}(K'r),\\
	\ttt_2(K,K',r)&=\frac{(2\pi)^{d}}{r^{d-2}(KK')^{\frac d2-1}}\Bigl(   \frac{d-1}{d}J_{\frac d2+1}(Kr)J_{\frac d2+1}(K'r)+\frac{1}{d}J_{\frac d2-1}(Kr)J_{\frac d2-1}(K'r) \Bigr),\\
	\ttt_3(K,K',r)&=\frac{(2\pi)^{d}}{r^{d-2}(KK')^{\frac d2-1}}\Bigl(   \frac{d-1}{d+2}J_{\frac d2+2}(Kr)J_{\frac d2+2}(K'r)+\frac{3}{d+2}J_{\frac d2}(Kr)J_{\frac d2}(K'r) \Bigr),
\end{align*}
whence
\begin{align}
	\ttt_{0,2}(K,K',r)&= (d-2)\ttt_0(K,K',r)+\ttt_2(K,K',r)\notag\\
	\hide{=\frac{(2\pi)^{d}}{r^{d-2}(KK')^{\frac d2-1}}
	\Bigl(   \frac{d-1}{d}J_{\frac d2+1}(Kr)J_{\frac d2+1}(K'r)+\frac{(d-1)^2}{d}J_{\frac d2-1}(Kr)J_{\frac d2-1}(K'r) \Bigr)\notag}
	&=\frac{d-1}{d}\frac{(2\pi)^{d}}{r^{d-2}(KK')^{\frac d2-1}}
	\Bigl(   J_{\frac d2+1}(Kr)J_{\frac d2+1}(K'r)+(d-1)J_{\frac d2-1}(Kr)J_{\frac d2-1}(K'r) \Bigr),\label{eqn:t02}\\
	\ttt_{1,3}(K,K',r)&= (d-3)\ttt_1(K,K',r)+2\ttt_3(K,K',r)\notag\\
	\hide{=\frac{(2\pi)^{d}}{r^{d-2}(KK')^{\frac d2-1}}\Bigl(   2\frac{d-1}{d+2}J_{\frac d2+2}(Kr)J_{\frac d2+2}(K'r)+\frac{d(d-1)}{d+2}J_{\frac d2}(Kr)J_{\frac d2}(K'r) \Bigr)\notag}
	&=\frac{d-1}{d+2}\frac{(2\pi)^{d}}{r^{d-2}(KK')^{\frac d2-1}}\Bigl(   2J_{\frac d2+2}(Kr)J_{\frac d2+2}(K'r)+d J_{\frac d2}(Kr)J_{\frac d2}(K'r) \Bigr)\label{eqn:t13}.
\end{align}

\section{Integrals of products of Bessel functions} 
\label{sec:integrals_of_bessel_functions}
We now record key integrals of products of Bessel functions. First, let's pull out two key formulae from Watson\cite{Watson}, namely Eqs.~(4) and (3) of \S 13.53:
\begin{align}
	\int_0^\infty\!\!\! \dd z \frac{z J_\mu(z)J_\mu(z)}{z^2-x^2} &= \frac{i\pi}2 H^{(1)}_{\mu}(x)J_{\mu}(x),\label{eqn:int Jmu Jmu sur z}\\
	\int_0^\infty\!\!\! \dd z \frac{zJ_\mu(z)J_{\mu+2}(z)}{z^2-x^2} &= \frac{i\pi}2 H^{(1)}_{\mu}(x)J_{\mu+2}(x).\label{eqn:int Jmu Jmu+2 sur z}
\end{align}
In extracting formulae out of Watson, one needs to exercise prudence with the treatment of subscripts. For instance, a careless reading of Watson's Eq.~(3) might suggest that $\int_0^\infty\!\!\! \dd z \frac{zJ_{\mu+2}(z)J_\mu(z)}{z^2-x^2} = \frac{i\pi}2 H^{(1)}_{\mu+2}(x)J_{\mu}(x)$.
Were this last equation true, however, we would then have $0=f(x)\equiv H^{(1)}_{\mu+2}(x)J_{\mu}(x)-H^{(1)}_{\mu}(x)J_{\mu+2}(x)$, while it is easy to prove that $f'(x)=-\frac 2x f(x)$, and  \killinpaper{thus $f(x)=icx^{-2}$ for some constant $c$. One can compute this constant by a careful analysis of the leading term. For $\mu>0$ and $\mu\not\in \mathbb{Z}$, we have $J_{\mu}(x)=\frac1{\Gamma(\mu+1)}\bigl(\frac x2\bigr)^\mu+\cdots$ while
$Y_\mu(x)=-\frac1{\Gamma(1-\mu)\sin(\mu\pi)}\bigl(\frac{x}2\bigr)^{-\mu}+\cdots$, thus
\begin{align*}
	H^{(1)}_{\mu+2}&(x)J_{\mu}(x)-H^{(1)}_{\mu}(x)J_{\mu+2}(x)\\ &=i(Y_{\mu+2}(x)J_{\mu}(x)-Y_{\mu}(x)J_{\mu+2}(x))\\
	&=-i\frac1{\Gamma(-\mu-1)\Gamma(1+\mu)\sin(\mu\pi)}\bigl(\frac x2\bigr)^{-2}+\cdots\\
	&=-i\frac{4(\mu+1)}{\pi}\frac1{x^2}+\cdots.
\end{align*} This last leading term is true even for $\mu\in\mathbb{Z}$. Thus}
\begin{equation}\label{eqn:switch HJ}
	H^{(1)}_{\mu+2}(x)J_{\mu}(x)-H^{(1)}_{\mu}(x)J_{\mu+2}(x)=-i\frac{4(\mu+1)}{\pi x^2}.
\end{equation}

To go from the first line of Eq.~\eqref{eqn:Master Eqn} to its contribution to Eq.~\eqref{eqn:B+- identities} only requires these two identities with $\mu=\frac d2\pm1$. To go from the second line requires  integrals of three Bessel functions. Such identities follow from the previous two. Simply substituting these previous results and using Eq.~\eqref{eqn:switch HJ} indeed yield
\begin{align}
	\int_0^\infty\!\!\!\dd z\frac{zJ_{\mu}(z)\Bigl(J_{\mu+2}(z)J_{\mu}(x)-J_\mu(z)J_{\mu+2}(x)\Bigr)}{z^2-x^2}
	\hide{=\frac{i\pi}2\Bigl(H^{(1)}_\mu(x)J_{\mu+2}(x)J_\mu(x)-H^{(1)}_\mu(x)J_\mu(x)J_{\mu+2}(x)\Bigr)\notag}
	&=0\label{eqn:int JJJ 0},\\
	\int_0^\infty\!\!\!\dd z\frac{zJ_{\mu+2}(z)\Bigl(J_{\mu+2}(z)J_{\mu}(x)-J_\mu(z)J_{\mu+2}(x)\Bigr)}{z^2-x^2}
	\hide{=\frac{i\pi}2\Bigl(H^{(1)}_{\mu+2}(x)J_{\mu+2}(x)J_{\mu(x)}-H^{(1)}_\mu(x)J_{\mu+2}(x)^2\Bigr)
	\notag}
\hide{=\frac{i\pi}2J_{\mu+2}(x)\Bigl(H^{(1)}_{\mu+2}(x)J_{\mu}(x)-H^{(1)}_\mu(x)J_{\mu+2}(x)\Bigr)\notag}
	\hide{=-i\frac{4(\mu+1)}{\pi}\frac{i\pi}2\frac{J_{\mu+2}(x)}{x^2}\notag}
	&=2(\mu+1)\frac{J_{\mu+2}(x)}{x^2}.\label{eqn:int JJJ not 0}
\end{align}

Using Eq.~\eqref{eqn:nuJnu} in concert with Eqs.~\eqref{eqn:int Jmu Jmu sur z} and \eqref{eqn:int Jmu Jmu+2 sur z}, we find
\begin{align}
	\int_0^\infty\!\!\!\dd z \frac{J_{\nu}^2(z)}{z(z^2- x^2)} 
	\hide{ = \frac1{4\nu^2}\int_0^\infty\!\!\!\dd z\frac{z\bigl(J_{\nu-1}(z)^2+2J_{\nu-1}(z)J_{\nu+1}(z)+J_{\nu+1}(z)^2\bigr)}{z^2- x^2}\notag}
	\hide{ = \frac{i\pi}{8\nu^2}\bigl(H^{(1)}_{\nu-1}( x)J_{\nu-1}( x)+2H^{(1)}_{\nu-1}( x)J_{\nu+1}( x)+H^{(1)}_{\nu+1}( x)J_{\nu+1}( x)\bigr)\notag}
	\hide{=\frac{i\pi}{8\nu^2}\Bigl(H^{(1)}_{\nu-1}( x)\bigl(J_{\nu-1}( x)+J_{\nu+1}( x)\bigr)+J_{\nu+1}( x)\bigl(H^{(1)}_{\nu-1}( x)+H^{(1)}_{\nu+1}( x)\bigr)\Bigr)\notag}
	&=\frac{i\pi}{4\nu x}\bigl(H^{(1)}_{\nu-1}( x)J_\nu( x)+J_{\nu+1}( x)H^{(1)}_{\nu}( x)\bigr).\label{eqn:int J2 over z z2-x2}
\end{align}
\killinpaper{It looks like the series of this might be equal to the series of $ x^{-2}(I_{\mu}(-i x)K_\mu(-i x)-\frac2\mu)$, at least for $\mu=\frac32$. Note the $-i x$ instead of $i x$. Note that it is tempting to use \S13.53 Eqn(6) of Ref.~\onlinecite{Watson} for this integral but this one would not give the $-\frac{2}{\nu x^2}$ term.}

\section{Integration over $[0,A]$}
\label{sec:integration_over_0_a}
In order to integrate over the finite interval $[0,A]$, we rely on the following crucial identities
\begin{align}
		\label{eqn:crucialIdentity}
	\int_0^A\!\!\!\! \dd r\, r J_\mu(K_1r)J_\mu(K_2r) &= \frac{A}{K_1^2-K_2^2}\Bigl(K_1J_{\mu+1}(A K_1)J_\mu(A K_2)-K_2J_{\mu+1}(A K_2)J_{\mu}(A K_1)\Bigr),\\
	\label{eqn:crucialIdentity2}
	\int_0^A\!\!\!\! \dd r\, r J_\mu(K_1r)J_\mu(K_2r)
	&=\frac{A}{K_1^2-K_2^2}\Bigl(K_2J_{\mu}(AK_1)J_{\mu-1}(A K_2)-K_1J_{\mu-1}(A K_1)J_{\mu}(A K_2)\Bigr).
\end{align}
\end{widetext}

\newcommand{\down}{\mathrm{down}}
\newcommand{\up}{\mathrm{up}}
\newcommand{\splitt}{\mathrm{split}}

We then define two operators that represent the above integration from $0$ to $A$:
\begin{enumerate}
	\item $\down_\mu$  replaces $r^{2-d}J_\mu(K_1r)J_\mu(K_2r)$ with the right-hand-side of Eq.~\eqref{eqn:crucialIdentity},
and leaves $rJ_\nu(Kr)J_\nu(K'r)$ intact for $\nu\neq\mu$
\item $\up_\mu$ replaces $r^{2-d}J_\mu(K_1r)J_\mu(K_2r)$ with the right-hand-side of Eq.~\eqref{eqn:crucialIdentity2}
and leaves $rJ_\nu(Kr)J_\nu(K'r)$ intact for $\nu\neq\mu$.
\end{enumerate}
It is important to emphasize that both $\down_\mu$ and $\up_\mu$ represent the integration from $0$ to $A$,  the result of which can be expressed in many different ways.   For instance, as $\ttt_{0,2}$ contains Bessel functions of order $\frac{d}2\pm1$, we have
\begin{align*}
\down_{\frac{d}2+1}\circ \down_{\frac d2-1}(\ttt_{0,2})&=\int_0^A\!\!\! \dd r r^{d-1} \ttt_{0,2}\\
& =  \up_{\frac{d}2+1}\circ \down_{\frac d2-1}(\ttt_{0,2}).	
\end{align*}
While expressed in terms of different Bessel functions, the quantities $\down_{\frac{d}2+1}\circ \down_{\frac d2-1}(\ttt_{0,2})$ and $\up_{\frac{d}2+1}\circ \down_{\frac d2-1}(\ttt_{0,2})$ are of course equal.

To clean up the resulting expression, we use 
\begin{equation}\label{eqn:nuJnu}
	J_{\nu-1}(x)+J_{\nu+1}(x)=\frac{2\nu}{x}J_\nu(x),
\end{equation}
and its two incarnations:
\begin{enumerate}
\item $\splitt_\nu$ replaces all occurences of $J_\nu(x)$ by $\frac{x}{2\nu}(J_{\nu-1}(x)+J_{\nu+1}(x))$, no matter what $x$ is, and leaves all the $J_\mu(x)$ intact for $\mu\neq\nu$;
\item $\splitt^-_\nu$ replaces occurences of $J_\nu(x)$ by 
$\frac{2(\nu-1)}{x}J_{\nu-1}(x)-J_{\nu-2}(x)$, no matter what $x$ is, and leaves all the $J_\mu(x)$ intact for $\mu\neq\nu$.
\end{enumerate}

We can thus compute $\tilde\RRi_A(K,K')$ using the sequence
\[\down_{\frac d2+2}\circ\down_{\frac d2+1}\circ\down_{\frac d2}\circ\up_{\frac d2-1}.\]
One then applies $\splitt_{\frac d2}\circ \splitt^-_{\frac d2+2}$ to obtain a remarkably simple expression with Bessel functions with only $\frac d2\pm1$ indices, namely Eq.~\eqref{eqn:SA in text}.  

\end{document}